\documentclass[11pt]{article}

\usepackage{a4wide}
\usepackage{cite}

\usepackage{amsmath}
\usepackage{amstext,amsfonts,amsbsy,eucal,amssymb}

\numberwithin{equation}{section}

\newtheorem{theorem}{Theorem}[section]
\newtheorem{lemma}[theorem]{Lemma}
\newtheorem{corollary}[theorem]{Corollary}
\newtheorem{conjecture}[theorem]{Conjecture}
\newtheorem{proposition}[theorem]{Proposition}

\newtheorem{problem}[theorem]{Problem}

\def\openone{\leavevmode\hbox{\small1\kern-3.8pt\normalsize1}}

\DeclareMathOperator{\Det}{Det}
\DeclareMathOperator{\Tr}{Tr}
\DeclareMathOperator{\diag}{diag}

\DeclareMathOperator{\ord}{ord}

\begin{document}

\baselineskip 21pt
\parskip 7pt

%\hfill{September 11, 2003}
%\\[10mm]
%%%%%%%%%%%%%%%%%% TITLE %%%%%%%%%%%%%%
%
\noindent
  {\LARGE{
      The matrix realization of affine Jacobi varieties \\[3mm]
      and the extended Lotka-Volterra lattice  
      }
    }
\\[5mm]
%
%%%%%%%%%%%%%%%%%%%%%%% AUTHOR(S) %%%%%%%%%%%%%%%%%%%
%
\begin{large}
Rei \textsc{Inoue}
  \footnote[2]{E-mail:
    \texttt{rei@gokutan.c.u-tokyo.ac.jp}
    }
\end{large}
\\[3mm]
%
%%%%%%%%%%%%%%%%%%%%%%% ADDRESS %%%%%%%%%%%%%%%%%%%%
%
  \textsl{Institute of Physics, Graduate School of Arts and Sciences,
    University of Tokyo \\ 
    Komaba 3--8--1, Meguro, Tokyo 153-8902, Japan.
    }
\\[3mm]
%
%%%%%%%%%%%%%%%%%%%%%% DATE %%%%%%%%%%%%%%%%%%%%%%%%
%
\baselineskip 15pt
\parskip 7pt
%
%%%%%%%%%%%%%%%%%%%%%% ABSTRACT %%%%%%%%%%%%%%%%%%%%%%
%
\begin{small}
\textbf{Abstract:} ~
We study completely integrable Hamiltonian systems 
whose monodromy matrices are related to 
the representatives for 
the set of gauge equivalence classes $\boldsymbol{\mathcal{M}}_F$
of polynomial matrices. 
Let $X$ be the algebraic curve given by 
the common characteristic equation for
$\boldsymbol{\mathcal{M}}_F$. 
We construct the isomorphism from
the set of representatives to
an affine part of the Jacobi variety of $X$.
This variety corresponds to the invariant manifold of the system,
where the Hamiltonian flow is linearized.
As the application, we discuss the
algebraic completely integrability of 
the extended Lotka-Volterra lattice with a periodic boundary condition.
\\[5mm]
%
%%%%%%%%%%%%%%%%%%%%%% Key Words %%%%%%%%%%%%%%%%%%%%%
%
%\textbf{Key Words:} ~
%
\end{small}

\baselineskip 15pt
\parskip 2pt

%%%%%%%%%%%%%%%%%%%%%%%%%%%%%%%%%%%%%%%%%%%%%%%
%%%%%%%%%%%%%%%%%%%%%%%%%%%%%%%%%%%%%%%%%%%%%%%

\section{Introduction}

The algebro-geometric structure of the completely integrable 
Hamiltonian systems was unveiled around 1980
(see \cite{DubMatNov76,Krichever78,MoerMum79,AdlerMoer80,ReySemenov94}
and references therein), and has been extensively studied. 
It was a remarkable discovery that
the Hamiltonian flows of the systems are linearized on 
 algebraic varieties 
like the Jacobi variety $J(X)$ of an algebraic curve $X$.
Many of the systems are 
described by the Lax equation of (Laurent) polynomial matrices of 
a spectral parameter,
and $X$ comes from its fixed characteristic equation
which gives the level set of the Lax matrix.
Typically the flows are linearized by the following procedure;
\begin{align}
  \label{procedure}
  \begin{split}
  \text{System} 
  &~~ \stackrel{\text{(I)}}{\longrightarrow} ~~ \text{Lax matrix} ~( \rightarrow ~ X)
  \\
  &~~ \stackrel{\text{(II)}}{\longrightarrow} ~~ \text{Div}_{\text{eff}}(X)
  \\
  &~~ \stackrel{\text{(III)}}{\longrightarrow} ~~ J(X)
  \end{split}
\end{align}
where Div$_{\text{eff}}(X)$ is the set of effective divisors.
The arrows (II) and (III) are respectively induced by the eigenvector map
and the Abel map.
In many cases, the linearization of the flows are related to the 
Lie algebraic symmetry of Lax matrices \cite{ReySemenov94}.
On the other hand, in \cite{Griffiths85}
the condition of the linearization was discussed 
based on a cohomological interpretation of the Lax equation.

Roughly speaking, 
(I) is heuristic, and (II) and (III) are systematic.
By the Abel-Jacobi theorem (III) is understood in general framework,
but (II) depends on the Lax matrix very much.  
Beauville showed that if we replace the Lax matrix with 
a set of gauge equivalence classes of 
polynomial matrices, (II) becomes an isomorphism \cite{Beauville90}.
He further proved that over the tangent space of the set
there exists the $g$ dimensional invariant vector field 
linearized on $J(X)$, where $g$ is the genus of $X$.
As claimed in \cite{Smirnov-Zeitlin0203},
to study concrete integrable systems 
we need to choose the orbit 
which gives the representative of the gauge equivalence class.
In fact, 
Mumford already gave an important example
when $X$ was a hyperelliptic curve \cite{Mumford-Book},
and introduced the set of representatives
with the explicit isomorphic maps (II) and (III). 
The dynamical system he introduced is called Mumford system, and 
has been studied from many points of view
\cite{DonagiMarkman96,Medan99,SmirnovNakayashiki00,Van1638,Vivolo03}.
Recently Smirnov and Zeitlin constructed the representative 
of the wider class of gauge equivalence classes,
by starting with $N$ by $N$
monodromy matrices of some special forms
\cite{Smirnov-Zeitlin0111,Smirnov-Zeitlin0203}.
They constructed the isomorphism (II) by making use of the  
separation of variables (SoV) a l\'a Sklyanin \cite{Sklyanin95}.

In this paper,
we consider the extension of \cite{Smirnov-Zeitlin0111},
and construct the isomorphic map (II) for a certain class of 
monodromy matrices.
We introduce $N$ by $N$ monodromy matrices 
$\mathbf{T}_{m;n_1,n_2}(z)$
($n_1 = 1,\cdots, N-1,~ n_2 = 1, \cdots, N$),
whose entries are
polynomials of a spectral parameter $z$ of degree $m$.
We fix a level set of $\mathbf{T}_{m;n_1,n_2}(z)$,
where the characteristic polynomial of $\mathbf{T}_{m;n_1,n_2}(z)$
is fixed to be $F_{m;n_1,n_2}(z,w) \in \mathbb{C}[z,w]$.
We write this set as $\{\mathbf{T}_{m;n_1,n_2}(z)\}_{F_{m;n_1,n_2}}$. 
The characteristic equation $F_{m;n_1,n_2}(z,w) = 0$
defines the complete algebraic curve $X$ and  
the set of gauge equivalence classes
$\boldsymbol{\mathcal{M}}_{F_{m;n_1,n_2}}$.
Let $\{\mathbf{M}(z)\}_{F_{m;n_1,n_2}}$ be the set of representatives
of $\boldsymbol{\mathcal{M}}_{F_{m;n_1,n_2}}$. 
Starting with the level set $\{\mathbf{T}_{m;n_1,n_2}(z)\}_{F_{m;n_1,n_2}}$, 
we study the following diagram
\begin{align}
  \label{eigenvector-map}
  \begin{split}
  &\{\mathbf{T}_{m;n_1,n_2}(z)\}_{F_{m;n_1,n_2}} 
   ~ \stackrel{\text{(b)}}{\longrightarrow}~
   X(g) 
  \\
  &~~~~~~~~~~~~~~~~~~ \text{\scriptsize (a)} \downarrow \hspace{1.3cm} 
   \nearrow \text{\scriptsize (c)}
  \\
  &\hspace{2.5cm}\{\mathbf{M}(z)\}_{F_{m;n_1,n_2}}
  \end{split}  
\end{align}
where $X(g) \subset \text{Div}_{\text{eff}}(X)$ 
is the set of effective divisors of degree $g$.
The map (a) is the gauge transformation, (b) is based on SoV,
and we construct these two so as to make the diagram 
\eqref{eigenvector-map} commutative.
Then the map (c) produces nothing but the case that 
(II) is isomorphic.
In \cite{Smirnov-Zeitlin0111}, the maps in \eqref{eigenvector-map} 
were given for $\{\mathbf{T}_{m;1,1}(z)\}_{F_{m;1,1}}$ of general $N$.
We study \eqref{eigenvector-map} in detail
for general cases of $N=2$ and $3$ here.

Next, as an application we study the integrable Hamiltonian structure of 
the extended Lotka-Volterra lattice. 
This is defined by the differential-difference equation
\begin{equation}
  \label{Bogo}
  \frac{\mathrm{d} V_n}{\mathrm{d} t}
  =
  2 \,V_n \,
  \sum_{k=1}^{N-1} \left(V_{n+k} - V_{n-k} \right),
\end{equation}
where $V_n \equiv V_n(t) \in \mathbb{C}$ for $n \in \mathbb{Z}$.
This model has the Hamiltonian structure and 
a family of integrals of motion in involution
\cite{Bogo88,Suris94,InoueHikami98-Bogo}.
When the model is infinite dimensional,
the $N=2$ case is known as the lattice KdV hierarchy 
\cite{FadTak86},
and the general $N>2$ case is related to the lattice $N$-reduced KP hierarchy
\cite{BonoraColCon96,AntoBelovChal97,FrenkelReshSemenov97,HikamiSogoInoue97}.
We set a periodic boundary condition $V_{n+L} = V_n$
for $L \in \mathbb{Z}_{\geq 2N-1}$,
and write LV($N,L$) for this finite dimensional model.
In \cite{Vanhae01}
the algebraic completely integrability of LV($2,L$)
was shown based on the analogues of the Mumford system, and
its invariant manifold is associated with an 
affine part of the Prym variety. 
Now, as the sequel of \cite{Inoue02-JBogo}
we study the integrability of LV($N,L$)  
by applying the structure \eqref{eigenvector-map}. 
We show that   
the monodromy matrix of LV($N,L$) is related to $\mathbf{T}_{m;n_1,n_2}(z)$
where the correspondence $L \leftrightarrow (m,n_1,n_2)$ is 
determined uniquely,
and that
the Poisson structure over $\{\mathbf{M}(z)\}_{F_{m;n_1,n_2}}$
is nicely embedded in that of LV($N,L$).
These enable SoV to describe explicitly the map (c)
as algebraic relations 
between the divisors in $X(g)$ and the dynamical variables $V_n$'s.
Finally  we give another proof of the algebraic completely integrability
of $N=2$ case, and newly show the $N=3$ case;
\begin{theorem}
  \label{th:N=2-3}
  LV($N,L$) is algebraic completely integrable 
 for $L \in \mathbb{Z}_{\geq 2N-1}$,  $N=2$ and $3$. 
\end{theorem}
We believe that it is true for general $N$. 

This paper is arranged as follows;
in \S 2, 
after a preliminary
we introduce
a class of $N$ by $N$ monodromy matrices $\mathbf{T}_{m;n_1,n_2}(z)$
which satisfy the fundamental Poisson relation
with the classical $r$-matrix.
By starting with these matrices 
we explain how to construct the maps in \eqref{eigenvector-map}.
In \S 3, 
we study the $N=2,3$ cases, where
the set of representatives $\{\mathbf{M}(z)\}_{F_{m;n_1,n_2}}$ and 
the eigenvector map (c) \eqref{eigenvector-map} are explicitly obtained.
In \S 4, 
we discuss the Hamiltonian structure of LV($N,L$)
and prove Theorem \ref{th:N=2-3}. 

The advantage of our way to investigate LV($N,L$)
is that we obtain the isomorphic eigenvector map
explicitly written as algebraic relations
between the divisor and the dynamical variables.
On the other hand, as discussed in \cite{Vanhae01},
for a model given by homogeneous evolution equations like
\eqref{Bogo},  
the Painlev\'e analysis \cite{AdlerMoer89} becomes a powerful tool
to construct the associated invariant manifold.
It may be interesting to study
the invariant manifold for LV($N,L$) 
based on these two viewpoints.

%%%%%%%%%%%%%%%%%%%%%%%%%%%%%%%%%%%%%%%%%%%%
%%%%%%%%%%%%%%%%%%%%%%%%%%%%%%%%%%%%%%%%%%%%

\section{Representatives for $\boldsymbol{\mathcal{M}}_F$ and eigenvector map}

\subsection{Preliminary}

Fix a polynomial $F(z,w)$ of the form
\begin{align}
  \label{curve-general}
  F(z,w)
  \equiv
  w^N - f_1(z) w^{N-1} + f_2(z) w^{N-2} - \cdots
  + (-1)^N f_N(z),
\end{align}
where each polynomial $f_i(z)$ satisfies $\text{deg} f_i(z) \leq i m$.
Let $X$ be the complete algebraic curve defined by 
$F(z,w) = 0$.
We assume $X$ is smooth, and let $g$ be its genus.
Let $\boldsymbol{\mathcal{M}}_F$ be the set of gauge equivalence classes
of $N$ by $N$ matrices whose matrix elements are
polynomials of $z$ of degree $m \in \mathbb{Z}_{>0}$;
\begin{align}
  \label{gauge-equiv}
  \boldsymbol{\mathcal{M}}_F
   =
  \{ \mathbf{M}(z) ~|~
      &\deg(\mathbf{M}(z)_{i,j}) \leq m \text{ for all } i,j, ~
      \nonumber \\ 
      &\Det \bigl( w \openone - \mathbf{M}(z) \bigr) = F(z,w) \}
   ~/~ \mathbf{GL}_N(\mathbb{C}).
\end{align}

For $\boldsymbol{\mathcal{M}}_F$
Beauville introduced the isomorphism \cite{Beauville90}
\begin{equation}
  \label{M-Div}
  \boldsymbol{\mathcal{M}}_F  \simeq X(g) - D.
\end{equation}
Here
$X(g)$ is 
the set of effective divisors $X(g) = X^g / \mathfrak{S}_g \subset
\mathrm{Div}_{\mathrm{eff}}(X)$, 
$\mathfrak{S}_g$ is the symmetric group 
and $D$ is a subset of $X(g)$.
The Abel map induces the isomorphism,
\begin{equation}
  \label{Div-Jac}
  X(g) - D \simeq J(X) - \Theta,
\end{equation}
where $D$ is mapped to a $(g-1)$-dimensional subvariety $\Theta$
called the theta divisor 
of the Jacobi variety $J(X)$.

We call $J(X) - \Theta$ the affine Jacobi variety of $X$ and
write $J_{\text{aff}}(X)$ for it.
We denote the set of representatives of 
$\boldsymbol{\mathcal{M}}_F$ using $\{\mathbf{M}(z)\}_F$.
Due to \eqref{M-Div} and \eqref{Div-Jac}
$\{\mathbf{M}(z)\}_F$ gives the matrix realization 
of $J_{\text{aff}}(X)$.
Herewith the arrows (II) and (III) in the procedure \eqref{procedure} 
becomes isomorphisms (II$^\prime$) and (III$^\prime$);
\begin{align}
  \label{iso-procedure}
  \begin{split}
  \{\mathbf{M}(z)\}_F
  ~~ &\stackrel{\text{(II$^\prime$)}}{\longrightarrow} 
  ~~ X(g) - D
  \\
  &\stackrel{\text{(III$^\prime$)}}{\longrightarrow} 
  ~~ J_{\text{aff}}(X).
  \end{split}
\end{align}

In this article, we let $\mathbf{M}_N(\mathbb{C})$ be a set of $N$ by $N$
complex matrices, $\mathbf{E}_{i,j}$ be an $N$ by $N$ basic matrix
as $(\mathbf{E}_{i,j})_{m,n} = \delta_{m,i}\delta_{n,j}$,
and $\vec{e}_i$ be an $N$ dimensional low vector
whose entries are zero but $i$-th is one.

%%%%%%%%%%%%%%%%%%%%%%%

\subsection{Classification of monodromy matrices and 
$\boldsymbol{\mathcal{M}}_F$}

We introduce lower/upper triangular $N$ by $N$ matrices,
\begin{align}
  \label{mu-}
  \begin{split}
  &\boldsymbol{\mu}^{(i)}_- 
   =
   \text{\tiny $i+1 \rightarrow$} 
   \begin{pmatrix}
     0 & \cdots & & & & & \cdots & 0\\
     \vdots & & & & & & & \vdots\\
     0 & \cdots & & & & & \cdots & 0\\
     \ast & \ast & 0 & \cdots & & & \cdots & 0 \\
     \ast & \ast & \ast & 0 & \cdots & & \cdots & 0\\
     \vdots & & & \ddots & \ddots & & & \vdots\\
     \ast & \cdots & \cdots & \cdots&  \ast & 0 & \cdots & 0 
   \end{pmatrix}, \text{ for } i= 1,\cdots, N-1, 
   \\ &\hspace*{5.7cm} \text{\tiny $\uparrow N+1-i$}
   \\
   & \hspace*{3.6cm} \text{\tiny $\downarrow i$}
   \\
   &\boldsymbol{\mu}^{(i)}_+
   =
   \begin{pmatrix}
     0 & \cdots & 0 & \ast & \cdots & \cdots & \ast \\
     \vdots & & & \ddots & \ast & \cdots & \ast\\
     \vdots & & & & \ddots  & \ddots & \vdots \\
     0 & \cdots & & & \cdots & 0 & \ast \\  
     0 & \cdots & & & & \cdots & 0 \\  
     \vdots & & & & & & \vdots \\ 
     0 & \cdots & & & & \cdots & 0 \\   
   \end{pmatrix} \text{\tiny $\leftarrow N+1-i$}~, 
   ~\text{ for } i= 1,\cdots, N,
  \end{split}
\end{align}
where {\tiny $i \rightarrow$} (or {\tiny $\downarrow i$}) indicates 
the $i$-th low (or column)
of the matrices, and  $\ast$ denote non-zero entries
which will be constants or variables.
For $N \geq 3$ we also use 
\begin{align}
   \label{mu-0}
   \begin{split}
   & \hspace*{3.3cm} \text{\tiny $\downarrow i+2$}
   \\
   &\boldsymbol{\mu}^{(-i)}_- 
   =
   \begin{pmatrix}
     \ast & \cdots & \ast & 0 & \cdots & 0 \\
     \ast & \cdots & \cdots & \ast & \ddots & \vdots \\
     \vdots & & & & \ddots & 0\\
     \ast & \cdots & & & \cdots & \ast \\
     \ast & \cdots & & & \cdots & \ast \\
     \vdots & & & & & \vdots \\
     \ast & \cdots & \cdots & \cdots & \cdots & \ast \\   
   \end{pmatrix}
   {\text{\tiny $\leftarrow N-1-i$}},
   ~~ \text{for $i=0,\cdots N-3$},
   \\
   &\boldsymbol{\mu}^{(-i)}_+ 
   =
   \text{\tiny $i+2 \rightarrow$}
   \begin{pmatrix}
     \ast & \cdots & \cdots & \cdots &\cdots & \ast \\[1mm]
     \vdots & & & & & \vdots \\
     \ast & \cdots &   & & \cdots & \ast \\[1mm]
     \ast & \cdots &   & & \cdots & \ast \\
     0 & \ddots & & & & \vdots\\
     \vdots & \ddots &\ast & \cdots & \cdots & \ast \\
     0 & \cdots & 0 & \ast & \cdots & \ast \\   
   \end{pmatrix}, ~~ \text{for $i=0,\cdots N-3$}.
   \\
   &\hspace{5.1cm}\text{\tiny $\uparrow N-i-1$}
   \end{split}
\end{align}
Using $\boldsymbol{\mu}_j, ~ j \in \mathbb{Z}_{>0}$,
we denote $N$ by $N$ matrices whose entries are 
not identically zero. 
We write $(\boldsymbol{\mu}_-^{(i)} \cap \boldsymbol{\mu}_+^{(j)})$
for a matrix which has zero at $(j_1,j_2)$ 
if $(\boldsymbol{\mu}_-^{(i)})_{j_1,j_2}$ or 
$(\boldsymbol{\mu}_+^{(j)})_{j_1,j_2}$ is zero. 
Note that 
$(\boldsymbol{\mu}_-^{(i)} \cap \boldsymbol{\mu}_j)$ and 
$\boldsymbol{\mu}_-^{(i)}$ have the same form.

First we fix the matrices \eqref{mu-}, \eqref{mu-0}
and $\boldsymbol{\mu}_j$ for $j = 1,\cdots,m-1$
to be constant matrices in $\mathbf{M}_N(\mathbb{C})$ as
$\boldsymbol{\mu}^{(i)}_- \equiv \boldsymbol{\mu}^{(i) 0}_-,  
\boldsymbol{\mu}^{(i)}_+ \equiv \boldsymbol{\mu}^{(i) 0}_+$,
and 
$\boldsymbol{\mu}_j \equiv \boldsymbol{\mu}_j^0$.
Using these matrices we define a set of $N$ by $N$ polynomial matrices 
of the spectral parameter $z \in \mathbb{C}$; 
\begin{equation}
  \label{general-T}
  \boldsymbol{\mathcal{T}}_{N}(z) 
  = 
  \bigl\{ \mathbf{T}_{m;n_1,n_2}^0(z) ~|~  
  m \in \mathbb{Z}_{>0}, ~
  n_1 \in \{1,2,\cdots, N-1\}, ~n_2 \in \{1,2,\cdots, N\} \bigr\},
\end{equation} 
where $\mathbf{T}_{m;n_1,n_2}^0(z)$ are defined as 
\begin{align}
  \label{T0-general-form}
  &\mathbf{T}_{m;n_1,n_2}^0(z)
   = 
  \begin{cases}
  &\boldsymbol{\mu}_-^{(n_1) 0} z^m + 
    \boldsymbol{\mu}_-^{(n_1-N+1) 0} z^{m-1}
     + \boldsymbol{\mu}_2^0 z^{m-2} + \cdots
     + \boldsymbol{\mu}_{m-2}^0 z^2  
     \\[1mm]
    &\hspace{4.8cm} 
     + \boldsymbol{\mu}_+^{(n_2-N) 0} z + \boldsymbol{\mu}_+^{(n_2) 0},
   \text{ for $m \geq 3$},
  \\[2mm]
  &\boldsymbol{\mu}_-^{(n_1) 0} z^2 +
    (\boldsymbol{\mu}_-^{(n_1-N+1) 0} \cap \boldsymbol{\mu}_+^{(n_2-N) 0}) \, z
   + \boldsymbol{\mu}_+^{(n_2) 0}, 
  \text{ for $m=2$},
  \\[2mm]
  &(\boldsymbol{\mu}_-^{(n_1) 0} \cap \boldsymbol{\mu}_+^{(n_2-N) 0}) \, z 
   + (\boldsymbol{\mu}_-^{(n_1-N+1) 0} \cap \boldsymbol{\mu}_+^{(n_2) 0}),
   \text{ for $m=1$}.
  \end{cases}
\end{align}
When $\boldsymbol{\mu}_-^{(n_1-N+1)}$ (or $\boldsymbol{\mu}_+^{(n_2-N)}$)
is not defined by \eqref{mu-0},
set $\boldsymbol{\mu}_-^{(n_1-N+1)} \equiv \boldsymbol{\mu}_1^0$
(or $\boldsymbol{\mu}_+^{(n_2-N)} \equiv \boldsymbol{\mu}_{m-1}^0$).
 
\begin{proposition}
  \label{T-F-injection}
  The map 
  \begin{align}
    \label{curve-F}
    \boldsymbol{\mathcal{T}}_{N}(z) &\rightarrow \mathbb{C}[z,w]; ~
    \mathbf{T}_{m;n_1,n_2}^0(z) \mapsto 
    F_{m;n_1,n_2}(z,w) 
    = 
    \Det \bigl( w \openone - \mathbf{T}_{m;n_1,n_2}^0(z) \bigr)
  \end{align}
  is injective.
\end{proposition}
{\em Proof.}
 Since the polynomial $F_{m;n_1,n_2}(z,w)$ \eqref{curve-F} 
 has a form as \eqref{curve-general},
 it is sufficient to check that 
 $\mathbf{T}_{m;n_1,n_2}^0(z) \mapsto  f_{N-1}(z)$ 
 is injective.
 Notice
 $$ f_{N-1}(z) = \Det \mathbf{T}_{m;n_1,n_2}^0(z) \cdot 
              \Tr (\mathbf{T}_{m;n_1,n_2}^0(z)^{-1}),
 $$
 and the forms of $\boldsymbol{\mu}_+^{(n_1)}$ and 
 $\boldsymbol{\mu}_-^{(n_2)}$ which compose $\mathbf{T}_{m;n_1,n_2}^0(z)$. 
 Then one sees 
 $$
   \deg f_{N-1}(z) = (N-1)m - n_1 + 1, ~~~
   \ord_{z=0} f_{N-1}(z) = n_2-1.
 $$ 
 Since $n_1 \in \{1, \cdots, N-1 \}$, $f_{N-1}(z)$ is classified by
 a triple $(m,n_1,n_2)$.
 In conclusion our claim is approved. $\square$
\\[2mm]
Therefore we see that 
$\mathbf{T}_{m;n_1,n_2}^0(z) \in \boldsymbol{\mathcal{T}}_{N}(z)$ 
corresponds to
$\boldsymbol{\mathcal{M}}_{F_{m;n_1,n_2}}$ 
\eqref{gauge-equiv} injectively.

Next we set
the entries of matrices \eqref{mu-}, \eqref{mu-0}
and $\boldsymbol{\mu}_j$ for $j = 1,\cdots,m-1$
to be variables,
and define $N$ by $N$ monodromy matrices $\mathbf{T}_{m;n_1,n_2}(z)$ 
$(m \in \mathbb{Z}_{>0}, ~ n_1 \in \{1,2,\cdots, N-1\}, 
~n_2 \in \{1,2,\cdots, N\})$ 
as same as \eqref{T0-general-form};
\begin{align}
  \label{T-general-form}
  &\mathbf{T}_{m;n_1,n_2}(z)
   = 
  \begin{cases}
  &\boldsymbol{\mu}_-^{(n_1)} z^m + 
    \boldsymbol{\mu}_-^{(n_1-N+1)} z^{m-1}
     + \boldsymbol{\mu}_2 z^{m-2} + \cdots + \boldsymbol{\mu}_{m-2} z^2 
     \\[1mm]
    &\hspace{4.5cm} 
     + \boldsymbol{\mu}_+^{(n_2-N)} z + \boldsymbol{\mu}_+^{(n_2)},
   \text{ for $m \geq 3$},
  \\[2mm]
  &\boldsymbol{\mu}_-^{(n_1)} z^2 +
    (\boldsymbol{\mu}_-^{(n_1-N+1)} \cap \boldsymbol{\mu}_+^{(n_2-N)}) \,z
   + \boldsymbol{\mu}_+^{(n_2)}, 
  \text{ for $m=2$},
  \\[2mm]
  &(\boldsymbol{\mu}_-^{(n_1)} \cap \boldsymbol{\mu}_+^{(n_2-N)}) \, z 
   + (\boldsymbol{\mu}_-^{(n_1-N+1)} \cap \boldsymbol{\mu}_+^{(n_2)}),
   \text{ for $m=1$}.
  \end{cases}
\end{align}
To study $\mathbf{T}_{m;n_1,n_2}(z)$, we define a local Lax matrix as
\begin{align}
  \label{general-Lax}
  \mathbf{L}_n(z) = \sum_{k=1}^{N-1} l_n^{(k)} \mathbf{E}_{k,k+1}
                      + z l_n^{(N)} \mathbf{E}_{N,1} 
                      + z l_n^{(0)} \mathbf{E}_{N,2},
\end{align}
where
$l_n^{(k)}$ $(n \in \mathbb{Z}, k = 0,\cdots, N$) are 
dynamical variables. 
\begin{lemma}
  \label{lemma-Lax}
  With the Lax matrix $\mathbf{L}_n(z)$ \eqref{general-Lax}
  the following Poisson relation is compatible;
 \begin{align}
  \label{L-r-poisson}
   \{ \mathbf{L}_n(z) \stackrel{\otimes}{,} \mathbf{L}_m(z^\prime) \}
   =
   \delta_{n,m}   [\, \mathbf{r}(z/z^{\prime}) ~,~ 
                  \mathbf{L}_n(z) \otimes \mathbf{L}_n(z^\prime) \,],
 \end{align}
where $\mathbf{r}(z)$ is the classical $r$-matrix 
\begin{align}
  \label{classical-r}
  &{\mathbf{r}}(z)
  = \frac{z + 1}{z - 1}
    \sum_{k=1}^N \mathbf{E}_{k,k} \otimes \mathbf{E}_{k,k}
      +
    \frac{2}{z-1}
    \sum_{1 \leq j < k \leq N}
    \Bigl( z \, \mathbf{E}_{k,j} \otimes \mathbf{E}_{j,k}
           + \, \mathbf{E}_{j,k} \otimes \mathbf{E}_{k,j}
    \Bigr).
\end{align} 
\end{lemma}
{\em Proof.}
  It is shown by a direct calculation.
  One easily sees that \eqref{L-r-poisson} is consistent
  with the Poisson bracket algebra for $l_n^{(k)} ~(k = 0,\cdots, N)$ defined 
  as
   \begin{align*}
    \begin{split}
   &\{ l_n^{(k)} ~,~ l_m^{(j)} \} = 0, 
   \text{ for } 1 \leq k,j \leq N,
   \\
   &\{ l_n^{(0)} ~,~ l_m^{(k)} \} = 
    \delta_{n,m} (\delta_{k,N} - \delta_{k,1}) l_n^{(0)} l_n^{(k)}.
    ~~~~~ \square
    \end{split}
  \end{align*}
\\
We define integers $m,m_1$ and $m_2$ by
\begin{align}
  \label{m-s}
  m = \Bigl[\frac{L}{N(N-1)}\Bigr], ~ 
  m_1 = \Bigl[\frac{L}{N-1}\Bigr], ~ m_2 = \Bigl[\frac{L}{N}\Bigr],
\end{align}
and determine $k, k_1$ and $k_2$ 
using 
\begin{align}
  \label{k-s}
  L=(N-1) m_1 + k_1 = N m_2 + k_2 = N(N-1)m + k.
\end{align}
\begin{lemma}
  \label{T-Lproduct}
  The monodromy matrix $\mathbf{T}_{m;n_1,n_2}(z)$ \eqref{T-general-form}
  can be written as a product of $L$ Lax matrix $\mathbf{L}_n(z)$ 
  \eqref{general-Lax};
  \begin{align}
    \label{LT-corresp}
    &z^{-m_2} \prod_{n=1}^{L} \mathbf{L}_n(z)
    = 
    \begin{cases}
      \mathbf{T}_{m;1,1}(z), ~\text{ for } k_1 = k_2 = 0,
      \\ 
      \mathbf{T}_{m+1;N-k_1,k_2+1}(z), ~\text{ for } k_1, k_2 \neq 0, ~ 0 \leq k_1 - k_2 \leq N-2
      \\
      \mathbf{T}_{m+2;N-k_1,k_2+1}(z), ~\text{ for } k_1 - k_2 \leq -1,
    \end{cases}    
  \end{align}
  where integers $m,m_2,k_1$ and $k_2$ are defined 
  in \eqref{m-s} and \eqref{k-s}.
\end{lemma}
See Appendix A for the proof.
Due to Lemmas \ref{lemma-Lax} and \ref{T-Lproduct}, 
it is straightforward to obtain the following proposition;
\begin{proposition}
  \label{T-r-Poisson}
With the matrix $\mathbf{T}_{m;n_1,n_2}(z)$
the fundamental Poisson relation is compatible;
\begin{align}
  \label{T-Poisson}
  \{\mathbf{T}_{m;n_1,n_2}(z)
  \stackrel{\otimes}{,}
  \mathbf{T}_{m;n_1,n_2}(z^\prime) \}
  =
  [\, \mathbf{r}(z/z^{\prime}) ~,~ \mathbf{T}_{m;n_1,n_2}(z) \otimes
  \mathbf{T}_{m;n_1,n_2}(z^\prime) \,].
\end{align}
\end{proposition}

Let $\mathcal{A}_C$ be the Poisson bracket algebra  
over the polynomial ring generated by 
the coefficients of entries in $\mathbf{T}_{m;n_1,n_2}(z)$,
whose defining relation is \eqref{T-Poisson}.
Then \eqref{T-Poisson} implies
\begin{proposition}  \cite{FadTak87}
  \label{prop:Poisson}
(i) The determinant of $\mathbf{T}_{m;n_1,n_2}(z)$ belongs to
 the center of $\mathcal{A}_{C}$;
$$
  \{ \mathbf{T}_{m;n_1,n_2}(z) ~,~ 
    \Det \mathbf{T}_{m;n_1,n_2}(z^\prime) \} = 0.
$$
(ii)  The coefficients of the characteristic polynomial 
  of $\mathbf{T}_{m;n_1,n_2}(z)$
  are Poisson commutative;
$$
 \{ \Det \bigl( w \openone - \mathbf{T}_{m;n_1,n_2}(z) \bigr)~,~ 
    \Det \bigl( w^\prime \openone - \mathbf{T}_{m;n_1,n_2}(z^\prime) \bigr)
   \} = 0.
$$   
\end{proposition}

Using \eqref{curve-F} 
we  define the level set of $\mathbf{T}_{m;n_1,n_2}(z)$ as
\begin{align*}
  \{\mathbf{T}_{m;n_1,n_2}(z)\}_{F_{m;n_1,n_2}}
   =
  \{\mathbf{T}_{m;n_1,n_2}(z) ~|~ 
      \Det ( w \openone - \mathbf{T}_{m;n_1,n_2}(z)) 
  = F_{m;n_1,n_2}(z,w)\}.
\end{align*}
Let $X$ be the complete algebraic curve determined by
$F_{m;n_1,n_2}(z,w)=0$, and its genus be $g$.
We consider the cases of $g \geq 1$. 
In general $\{\mathbf{T}_{m;n_1,n_2}(z)\}_{F_{m;n_1,n_2}}$ 
constitutes a variety whose dimension is greater than $g$.
Since the isomorphism \eqref{M-Div} implies that 
$\boldsymbol{\mathcal{M}}_{F_{m;n_1,n_2}}$ is a $g$ dimensional variety,
we state a problem to construct the map (a) 
\eqref{eigenvector-map} which gives 
the set of representatives  
$\{\mathbf{M}(z)\}_{F_{m;n_1,n_2}}$ as follows;
\begin{problem}
  \label{problem}
For $\mathbf{T}_{m;n_1,n_2}(z)$
find a gauge matrix 
$\mathbf{S}$ on $\mathbf{T}_{m;n_1,n_2}(z)$,
such that the set 
\begin{align}
  \label{M-rep}
  \begin{split}
  &\{ \mathbf{M}(z) \}_{F_{m;n_1,n_2}}
  = 
  \{ \mathbf{M}(z) = \mathbf{S} \,  \mathbf{T}_{m;n_1,n_2}(z) \,\mathbf{S}^{-1}
       ~|~ 
      \Det ( w \openone - \mathbf{M}(z)) = F_{m;n_1,n_2}(z,w)\}
  \end{split}
\end{align}
constitutes a $g$ dimensional variety.
\end{problem}
We note that the matrix
$\mathbf{M}(z)$ 
has the same degree as 
$\mathbf{T}_{m;n_1,n_2}(z)$ as a polynomial matrix,
and write it as
\begin{equation}
  \label{general-M}
  \mathbf{M}(z) = 
  \boldsymbol{\eta}_0 z^m + \boldsymbol{\eta}_{1} z^{m-1} + 
  \cdots + \boldsymbol{\eta}_{m-1} z + \boldsymbol{\eta}_m.
\end{equation}
Here the variable matrices $\boldsymbol{\eta}_i$ do not depend on $z$.
Once the above problem is solved,
the Poisson bracket algebra generated by 
the matrix elements of $\boldsymbol{\eta}_i$ \eqref{general-M} 
is induced by $\mathcal{A}_C$, and
we let $\mathcal{A}_{M}$ be this algebra.
Due to Proposition \ref{prop:Poisson}, the coefficients of 
$\Det \bigl( w \openone - \mathbf{T}_{m;n_1,n_2}(z) \bigr)$
constitute the
commuting subalgebra of $\mathcal{A}_{M}$.

In the following, without any notice we pay attention to an element
of $\boldsymbol{\mathcal{T}}_N(z)$ \eqref{general-T}, 
and abbreviate $F_{m;n_1,n_2}$ to $F$. 

%%%%%%%%%%%%%%%%%%

\subsection{Eigenvector map and SoV}

Following \cite{Smirnov-Zeitlin0111,Smirnov-Zeitlin0203}
we introduce the eigenvector map (b) \eqref{eigenvector-map}
by making use of SoV. %\cite{Sklyanin85,Sklyanin95}.
Sklyanin refined the technique invented to solve
the spectral problem of the quantum Toda lattice,
and introduced the method called SoV based on 
the $R$-matrix structure of the monodromy matrices 
(See \cite{Sklyanin85,Sklyanin95} and references therein).
The SoV for the monodromy matrices of $SL(N)$ symmetry
has been studied in detail.
The cases of $N=2$ and $3$ are done by Sklyanin himself
\cite{Sklyanin85,Sklyanin92},
and the extension to the general $N$ cases are clarified in 
\cite{Scott94,Gekhtman95}.

For classical systems 
this method derives the canonically conjugate variables 
from the poles of the eigenvector of the monodromy matrix.
We review this mechanism following \cite{Sklyanin95}.
Let $\mathbf{T}(z)$ be an $N$ by $N$
monodromy matrix which satisfies the fundamental Poisson 
relation as \eqref{T-Poisson}.
Then the eigenvector of $\mathbf{T}(z)$ called the Baker-Akhiezer function
is defined as
\begin{align*}
  \mathbf{T}(z) \vec\phi(z) = w \vec\phi(z),
  ~~~ \sum_{n=1}^N a_n(z) \phi_n(z) = 1,
\end{align*}
where $\vec\phi(z) = (\phi_1(z), \cdots, \phi_N(z))$,
and $w$ is the eigenvalue. 
The second equation is a normalization which uniquely determines $\vec\phi(z)$.
When $\vec\phi(z)$ has a pole at $z=z_i$, the residues 
$\vec\phi_i = (\phi_{1,i},\cdots, \phi_{N,i}) = \mathrm{res}_{z = z_i} 
\vec\phi(z)$
satisfy
\begin{align}
  \label{BA-function}
  \mathbf{T}(z_i) \vec\phi_i = w_i \vec\phi_i,
  ~~~ 
  \sum_{n=1}^N a_n(z_i) \phi_{n,i} = 0.
\end{align}
Then the condition to get non-zero vector $\vec\phi_i$ becomes
\begin{align}
  \label{SoV-det}
  \Det 
  \begin{pmatrix}
     a_1(z) & a_2(z) & \cdots & a_N(z) \\
     T(z)_{1,1}-w & T(z)_{1,2} & \cdots & T(z)_{1,N} \\
     \vdots \\
     T(z)_{j-1,1} & T(z)_{j-1,2} & \cdots & T(z)_{j-1,N} \\
     T(z)_{j+1,1} & T(z)_{j+1,2} & \cdots & T(z)_{j+1,N} \\
     \vdots \\
     T(z)_{N,1} & T(z)_{N,2} & \cdots & T(z)_{N,N}-w 
  \end{pmatrix}
  = 0, 
  ~~~ \text{for $j=1, \cdots N$},
\end{align}
where $T(z)_{i,j} = (\mathbf{T}(z))_{i,j}$.

In our case with the monodromy matrix $\mathbf{T}_{m;n_1,n_2}(z)$
\eqref{T-general-form},
some simple choices of the vector $\vec{a}(z) = (a_1(z), \cdots, a_N(z))$  
give SoV, and  \eqref{SoV-det} reduces to two equations on
$\mathbf{T}_{m;n_1,n_2}(z)$ as \cite{Sklyanin95}
\begin{align}
  \label{separation}
  B(z) = 0, ~~~ w = A(z).
\end{align}
Here $A(z) = A(\mathbf{T}_{m;n_1,n_2}(z))$ 
is a rational function of $z$ and 
$B(z) = B(\mathbf{T}_{m;n_1,n_2}(z))$ is a polynomial.
Accordingly the zero of $B(z)$, $z_i$
uniquely determines the eigenvalue $w_i= A(z_i)$.
The significant benefit of the fundamental Poisson relation 
\eqref{T-Poisson} is that 
the variables $(z_i, w_i)$ turn out to be canonically conjugate variables,
namely they fulfill the canonical Poisson brackets,
\begin{align*}
  \{ z_i ~,~ z_j \} = \{ w_i ~,~ w_j \} = 0,
  ~~~
  \{ z_i ~,~ w_j \} = 2 \, \delta_{i,j} z_i w_i.
\end{align*}
These variables are nothing but the {\it separated variables},
and the equation $B(z) = 0$ \eqref{separation} is called
the separation equation.

When we consider the level set $\{ \mathbf{T}_{m;n_1,n_2}(z) \}_F$,
each pair $(z_i,w_i)$ satisfies
$F(z_i, w_i) = 0$.
We expect that the separation equation has a following form,
\begin{equation}
  \label{separationB}
  B(z) = B_0 z^{f(n_1,n_2)}\prod_{i=1}^{g} (z - z_i),
  ~~~ f(n_1,n_2) \in \mathbb{Z}_{\geq 0},
\end{equation}
where $g$ is the genus of the algebraic curve $X$ given by 
$F(z,w) = 0$. 
There are certainly some different choices of the 
separation equations \eqref{separation} depending on the vector 
$\vec a(z)$.
To make the diagram \eqref{eigenvector-map} commutative,
we should choose the separation equation invariant 
under the gauge $\mathbf{S}$ \eqref{M-rep}.

To close this section,
we mention the subset $D$ 
which appeared in the isomorphism \eqref{Div-Jac}.
We assume $(z_i,w_i) \neq (z_j,w_j)$ for $i \neq j$,
and a set of the $g$ separated variables $(z_i,w_i)$
determines an effective divisor 
\begin{align}
  \label{div-P}
  P = \sum_{i=1}^g [(z_i,w_i)] \in X(g). 
\end{align}
Then the subset $D$ should be set as \cite{Smirnov-Zeitlin0111}
\begin{equation}
  \label{D}
  D = \{ P = \sum_{i=1}^g [(z_i, w_i)] ~|~ 
         \Det \bigl( h_i(z_j,w_j) \bigr)_{1 \leq i,j \leq g} = 0 \},
\end{equation}
where $h_i(z,w)$ are defined by
homomorphic one-forms $\sigma_i$ on $X$ \cite{Griffiths-Book},
\begin{equation}
  \label{oneform}
  \sigma_i(z,w) = \frac{h_i(z,w) \mathrm{d}z}{\frac{\partial}{\partial w} F(z,w)},
  ~~\text{ for }i = 1, \cdots, g. 
\end{equation}
We remark that 
the $g$ independent vector fields on a tangent space of  
$\boldsymbol{\mathcal{M}}_F$
are generated by the coefficients of $F(z,w)$ \eqref{curve-F}.
The fundamental Poisson relation \eqref{T-Poisson} ensures that
the evolution of the divisor $P$ generated by the vector fields is 
linearized on $J_{\text{aff}}(X)$.

%%%%%%%%%%%%%%%%%%%%%%%%%%%%%%%%
%%%%%%%%%%%%%%%%%%%%%%%%%%%%%%%%

\section{Study of concrete cases}

Starting with $\mathbf{T}_{m;n_1,n_2}(z)$,
we study the diagram \eqref{eigenvector-map}.
We construct 
the gauge matrix $\mathbf{S}$ \eqref{M-rep} 
which gives 
the set of representatives $\{\mathbf{M}(z)\}_F$, and 
the associated separation equation which 
makes the map (c) \eqref{eigenvector-map} well-defined.  
Then the isomorphic eigenvector map (II$^\prime$) \eqref{iso-procedure}
is induced by (c).
We explicitly discuss the cases of $N=2$ and $3$ with $g \geq 1$.
Further we recall $\{\mathbf{M}(z)\}_F$ associated with 
$\mathbf{T}_{m;1,1}(z)$ for general $N$ 
\cite{Smirnov-Zeitlin0111,Inoue02-JBogo}. 

%%%%%%%%%%%%%%%%%%%%%%%%%
 
\subsection{N=2 case}

We have matrices \eqref{mu-}
\begin{equation*}
%  \label{mu-2}
  \boldsymbol{\mu}_-^{(1)} =
   \begin{pmatrix}
       0 & 0 \\
       \ast & \ast   
   \end{pmatrix}, 
~~ \boldsymbol{\mu}_+^{(1)} =
   \begin{pmatrix}
       \ast & \ast \\
       0 & \ast   
   \end{pmatrix},
~~ \boldsymbol{\mu}_+^{(2)} =
   \begin{pmatrix}
       0 & \ast \\
       0 & 0   
   \end{pmatrix}.
\end{equation*}
Using them we introduce two matrices,
$\mathbf{T}_{m;1,1}(z)$ and $\mathbf{T}_{m;1,2}(z)$,
and derive the associated representatives;

\noindent
(i) $\mathbf{T}_{m;1,1}(z)$:
We have the matrix
\begin{align}
  \label{T-m11}
  \mathbf{T}_{m;1,1}(z) = 
  \boldsymbol{\mu}_-^{(1)} z^m + \boldsymbol{\mu}_1 z^{m-1} + \cdots 
  + \boldsymbol{\mu}_{m-1} z + \boldsymbol{\mu}_+^{(1)}, 
  ~~ \text{for} ~ m \geq 2.
\end{align}
The spectral curve $X$ is given by
\begin{align}
  \label{ch-2-even}
  \begin{split}
  &F(z,w) = \Det(w \openone - \mathbf{T}^0_{m;1,1}(z))
         = w^2 - w f_1(z) + f_2(z) = 0, ~~
  \\ 
  & ~~\text{where} ~\text{deg}f_1(z) = m, ~~\text{deg}f_2(z) = 2m-1, 
  \end{split}
\end{align}
and its genus is $g=m-1$.
The set $\{\mathbf{M}(z)\}_F$ \eqref{M-rep}
is obtained as the level set of 
\begin{align}
  \label{gauge-2-even}
  \mathbf{M}(z) =
  {\mathbf{S}} \,  \mathbf{T}_{m;1,1}(z) \,{\mathbf{S}}^{-1},
   ~~ 
   {\mathbf{S}} = 
   \begin{pmatrix}
     \vec{e}_1 \\
     \vec{e}_1 \boldsymbol{\mu}_1
   \end{pmatrix},
\end{align}
where $\mathbf{M}(z)$ has the form as 
$$
  \mathbf{M}(z) = \boldsymbol{\eta}_0 z^m + \cdots + \boldsymbol{\eta}_m, 
  ~~ \text{where}~
  \boldsymbol{\eta}_0 = 
  \begin{pmatrix}
    0 & 0 \\
    \ast & \ast
  \end{pmatrix},
  ~~
  \boldsymbol{\eta}_1 = 
  \begin{pmatrix}
    0 & 1 \\
    \ast & \ast
  \end{pmatrix},
$$
and other $\boldsymbol{\eta}_i$ are the matrices without zero entries.

\noindent
(ii) $\mathbf{T}_{m;1,2}(z)$: 
This is the case with the matrix as
\begin{align*}
  \mathbf{T}_{m;1,2}(z) = 
    \boldsymbol{\mu}_-^{(1)} z^{m} + \boldsymbol{\mu}_1 z^{m-1} 
    + \cdots + \boldsymbol{\mu}_{m-1} z + \boldsymbol{\mu}_+^{(2)},
    ~~ \text{for} ~ m \geq 2,
\end{align*}
and $X$ is determined by 
\begin{align}
  \label{ch-2-odd}
  \begin{split}
  &F(z,w) = \Det( w \openone - \mathbf{T}^0_{m;1,2}(z) )
         = w^2 - w z f_1^\prime(z) + z f_2^\prime(z) = 0, 
  \\
  &~ \text{where} ~
  \text{deg} f_1^\prime(z) = m-1, ~~ \text{deg} f_2^\prime(z) = 2m-2.
  \end{split}
\end{align}
The genus of $X$ is $m-1$.
By using the gauge matrix
\begin{align}
   \label{gauge-2-odd}
   {\mathbf{S}} = 
   \begin{pmatrix}
     \vec{e}_2 \boldsymbol{\mu}_-^{(1)} \\
     \vec{e}_2 
   \end{pmatrix},
\end{align}
we obtain $\mathbf{M}(z)$  \eqref{general-M} with
$$
  \boldsymbol{\eta}_0 = 
  \begin{pmatrix}
    \ast & 0 \\
    1 & 0
  \end{pmatrix},
  ~~
  \boldsymbol{\eta}_{m} = 
  \begin{pmatrix}
    0 & \ast  \\
    0 & 0
  \end{pmatrix},
$$
and the other $\boldsymbol{\eta}_i$ are the matrices with no zero-entries.

One sees that
both of $\{\mathbf{T}_{m;1,1}(z)\}_F$ and 
$\{\mathbf{T}_{m;1,2}(z)\}_F$ 
constitute the algebraic varieties of $m$ dimension 
which is equal to the genus of $X$.
For example, by the definition \eqref{T-m11} 
one sees that $\mathbf{T}_{m;1,1}(z)$ has $(4m+1)$ variables
to which the fixed characteristic equation \eqref{ch-2-even} gives $3m+1$ 
relations. 
Then we see $\{\mathbf{T}_{m;1,1}(z)\}_F$ constitutes the
$m$-dimensional algebraic variety. 
The gauge matrix $\mathbf{S}$ reduce $\{\mathbf{T}_{m;1,1}(z)\}_F$ 
by one dimension, and
$\{\mathbf{M}(z)\}_F$ becomes $m$ dimensional.
By choosing the vector $\vec{a}(z) = (a_1(z),a_2(z))$ \eqref{BA-function}
the separation equation \eqref{separation} is obtained as
\begin{align}
  \label{separation2*2}
  B(z) = 
  \begin{cases}
    {T}(z)_{1,2} = B_0 \displaystyle{\prod_{i=1}^{m-1} (z- z_i)}, 
                         ~ \vec{a}(z) = (1,0) \text{ for (i)},\\
    {T}(z)_{2,1} = B_0 \, z \displaystyle{\prod_{i=1}^{m-1} (z- z_i)}, 
                         ~ \vec{a}(z) = (0,1) \text{ for (ii)},\\
  \end{cases}
\end{align}
where $T(z)_{i,j} = (\mathbf{T}_{m;n_1,n_2}(z))_{i,j}$.
In both cases $B(z)$ generally
has $m-1$ non-zero zeros; $z_1, \cdots, z_{m-1}$,
and each of them gives an eigenvalue
$$
  w_i =
  \begin{cases}
    T(z_i)_{2,2}, \text{ for (i)},\\
    T(z_i)_{1,1}, \text{ for (ii)}.\\
  \end{cases}
$$
In the level set $\{\mathbf{T}_{m;n_1,n_2}(z)\}_F$
the points $(z_i,w_i)$ on $X$ determine the effective divisor over $X$,
$$
  P = \sum_{i=1}^{m-1} [(z_i, w_i)] \in X(g) - D.
$$ 
We remark that
this divisor is invariant under the gauge transformation 
induced by $\mathbf{S}$,
namely the solution of the separation equation does not change 
after replacing each $T(z)_{i,j}$ with a matrix element
of $\mathbf{M}(z)$; $M(z)_{i,j}$. 
In this $N=2$ case $X$ is linearly transformed to the hyperelliptic curve,
and we can easily see the structure of $D$ \cite{Mumford-Book}. 
On the curve $X$, we have two infinity points $\infty_\pm$
and $m-1$ homomorphic one forms \eqref{oneform}  
$$ 
  h_i(z,w) = z^{i-1}, ~~ \text{ for } i = 1, \cdots, m-1.
$$  
Then $D$ is written as \cite{Smirnov0001}
$$
  D = \{ P = \sum_{i = 1}^{m-1} [(z_i, w_i)] 
         ~|~ z_i = z_j \text{ for some } i \neq j,
           \text{ or } (z_i,w_i) = \infty_{\pm} \text{ for some } i \}.    
$$

%%%%%%%%%%%%%%%%%%%%%%%%%%%%%

\subsection{N=3 case}

The matrices \eqref{mu-} and \eqref{mu-0} 
are written as
\begin{align*}
  &\boldsymbol{\mu}_-^{(1)} = 
  \begin{pmatrix}
     0 & 0 & 0\\
     \ast & \ast & 0\\
     \ast & \ast & \ast 
  \end{pmatrix}, 
  ~~
 \boldsymbol{\mu}_-^{(2)} = 
  \begin{pmatrix}
      0 & 0 & 0\\
      0 & 0 & 0\\
     \ast & \ast & 0 
  \end{pmatrix}, 
  ~~
 \boldsymbol{\mu}_-^{(0)} = 
  \begin{pmatrix}
      \ast & \ast & 0\\
      \ast & \ast & \ast\\
     \ast & \ast & \ast 
  \end{pmatrix}, 
\\[2mm] 
  &\boldsymbol{\mu}_+^{(1)} = 
  \begin{pmatrix}
      \ast & \ast & \ast\\
      0 & \ast & \ast\\
      0 & 0 & \ast 
  \end{pmatrix},
  ~~
 \boldsymbol{\mu}_+^{(2)} = 
  \begin{pmatrix}
      0& \ast & \ast\\
      0 & 0 & \ast\\
      0 & 0 & 0
  \end{pmatrix},
  ~~
 \boldsymbol{\mu}_+^{(3)} = 
  \begin{pmatrix}
      0 & 0 & \ast\\
      0 & 0 & 0\\
      0 & 0 & 0
  \end{pmatrix},
  ~~
 \boldsymbol{\mu}_+^{(0)} = 
  \begin{pmatrix}
      \ast & \ast & \ast\\
      \ast & \ast & \ast\\
      0 & \ast & \ast 
  \end{pmatrix}.
\end{align*}
We study $6$ cases of $\mathbf{T}_{m;n_1,n_2}(z)$,
$n_1 = 1,2$ and $n_2 = 1,2,3$.
For each $\mathbf{T}_{m;n_1,n_2}(z)$ we enumerate
the forms of the spectral curve $F(z,w)$ given by
$\mathbf{T}_{m;n_1,n_2}^0(z)$,
the gauge matrix $\mathbf{S}$ and the matrix $\mathbf{M}(z)$
\eqref{M-rep}.
In the following, unless a concrete form is shown,
$\boldsymbol{\eta}_i$ \eqref{general-M}
denote the matrices without zero-entries.

\noindent
(i) $\mathbf{T}_{m;1,1}(z)$:
We start with the matrix
$$
\mathbf{T}_{m;1,1}(z) =
 z^m \boldsymbol{\mu}_-^{(1)} + z^{m-1} \boldsymbol{\mu}_1 
 + \cdots + z \boldsymbol{\mu}_{m-1} + \boldsymbol{\mu}_+^{(1)},
$$
whose characteristic polynomial is given by $\mathbf{T}_{m;1,1}^0(z)$
as
\begin{equation}
  \label{F-3m}
  F(z,w) = w^3 - f_1(z) w^2 + f_2(z) w - f_3(z),
\end{equation}
where $\deg f_1(z) = m, ~\deg f_2(z) = 2m$ and $\deg f_3(z) = 3m-1$.
The genus of the curve $X$ is $g = 3m-2$.
The gauge matrix
$$
  \mathbf{S} = 
  \begin{pmatrix}
    \vec{e}_1 \\
    \vec{e}_1 \boldsymbol{\mu}_1 \boldsymbol{\mu}_-^{(1)} \\
    \vec{e}_1 \boldsymbol{\mu}_1 
  \end{pmatrix}
$$  
introduces \eqref{general-M} of the form;
$$
  \mathbf{M}(z) =
  z^m \begin{pmatrix}
       0 & 0 & 0 \\ 
       \ast & \ast & \ast \\
       0 & 1 & 0
      \end{pmatrix}
   +
  z^{m-1} \begin{pmatrix}
       0 & 0 & 1 \\ 
       \ast & \ast & \ast \\
       \ast & \ast & \ast \\
      \end{pmatrix}
   + O(z^{m-2}).
$$

\noindent
(ii) $\mathbf{T}_{m;2,2}(z)$:
The matrix is 
$$
\mathbf{T}_{m;2,2}(z) =
 z^{m} \boldsymbol{\mu}_-^{(2)} + z^{m-1} \boldsymbol{\mu}_-^{(0)} 
 + z^{m-2} \boldsymbol{\mu}_2 + \cdots + z \boldsymbol{\mu}_{m-1} 
 + \boldsymbol{\mu}_+^{(2)},
 ~~ \text{for} ~ m \geq 2, 
$$
and $X$ is given by $\mathbf{T}_{m;2,2}^0(z)$, 
\begin{equation}
  \label{F-3m+1}
  F(z,w) = w^3 - z f_1^\prime(z) w^2 + z f_2^\prime(z) w - z f_3^\prime(z),
\end{equation}
where $\deg f_1^\prime(z) = m-2, ~\deg f_2^\prime(z) = 2m-2$ 
and $\deg f_3^\prime(z) = 3m-3$.
The genus is $g = 3m-3$. Due to the gauge matrix 
$$
  \mathbf{S} = 
  \begin{pmatrix}
    \vec{e}_3 \boldsymbol{\mu}_-^{(2)} \\
    \vec{e}_3 \boldsymbol{\mu}_-^{(2)} \boldsymbol{\mu}_+^{(2)} \\
    \vec{e}_3 
  \end{pmatrix}
$$  
\eqref{general-M} is obtained as
$$
  \mathbf{M}(z) =
  z^{m} \begin{pmatrix}
       0 & 0 & 0 \\ 
       \ast & 0 & 0 \\
       1 & 0 & 0
      \end{pmatrix}
   + \cdots +
  \begin{pmatrix}
       0 & 1 & 0 \\ 
       0 & 0 & \ast \\
       0 & 0 & 0 \\
      \end{pmatrix}.
$$

\noindent
(iii) $\mathbf{T}_{m;1,3}(z)$:
For the matrix
$$
\mathbf{T}_{m;1,3}(z) =
 z^{m} \boldsymbol{\mu}_-^{(1)} + z^{m-1} \boldsymbol{\mu}_1 + \cdots
 + z \boldsymbol{\mu}_+^{(0)} + \boldsymbol{\mu}_+^{(3)},
  ~~ \text{for} ~ m \geq 2, 
$$
$X$ is given by 
\begin{equation}
  \label{F-3m+2}
  F(z,w) = w^3 - z f_1^\prime(z) w^2 + z^2 f_2^\prime(z) w - z^2 f_3^\prime(z),
\end{equation}
where $\deg f_1^\prime(z) = m-1, ~\deg f_2^\prime(z) = 2m-2$ 
and $\deg f_3^\prime(z) = 3m-3$.
The genus of $X$ is $g = 3m-3$.
The gauge 
$$
  \mathbf{S} = 
  \begin{pmatrix}
    \vec{e}_1 \\
    \vec{e}_1 \boldsymbol{\mu}_+^{(3)} \boldsymbol{\mu}_+^{(0)} \\
    \vec{e}_1 \boldsymbol{\mu}_+^{(3)}
  \end{pmatrix}
$$  
gives \eqref{general-M}, 
$$
  \mathbf{M}(z) =
  z^{m} \begin{pmatrix}
       0 & 0 & 0 \\ 
       \ast & \ast & \ast \\
       \ast & \ast & \ast
      \end{pmatrix}
   + \cdots +
   z \begin{pmatrix}
       \ast & \ast & \ast \\ 
       \ast & \ast & \ast \\
       0 & 1 & 0 \\
      \end{pmatrix}
   + \begin{pmatrix}
       0 & 0 & 1 \\
       0 & 0 & 0 \\
       0 & 0 & 0 
     \end{pmatrix}.
$$

\noindent
(iv) $\mathbf{T}_{m;2,1}(z)$:
The matrix is
$$
\mathbf{T}_{m;2,1}(z) =
 z^{m} \boldsymbol{\mu}_-^{(2)} + z^{m-1} \boldsymbol{\mu}_-^{(0)} 
 + z^{m-2} \boldsymbol{\mu}_2 + \cdots + z \boldsymbol{\mu}_{m-1} 
 + \boldsymbol{\mu}_+^{(1)},
 ~~ \text{for} ~ m \geq 2, 
$$
and $X$ is given by the form \eqref{F-3m} with
$\deg f_1(z) = m-1, ~\deg f_2(z) = 2m-1$ and $\deg f_3(z) = 3m-2$.
The genus of $X$ is $g = 3m-3$.
The gauge matrix and the matrix \eqref{general-M} are
$$
  \mathbf{S} = 
  \begin{pmatrix}
    \vec{e}_3 \boldsymbol{\mu}_-^{(2)} \\
    \vec{e}_3 \boldsymbol{\mu}_-^{(2)} \boldsymbol{\mu}_+^{(1)} \\
    \vec{e}_3 
  \end{pmatrix},
  ~~~
  \mathbf{M}(z) =
  z^{m} \begin{pmatrix}
       0 & 0 & 0 \\ 
       \ast & 0 & 0 \\
       1 & 0 & 0
      \end{pmatrix}
   + \cdots +
  \begin{pmatrix}
       0 & 1 & 0 \\ 
       \ast & \ast & \ast \\
       0 & 0 & \ast \\
      \end{pmatrix}.
$$

\noindent
(v) $\mathbf{T}_{m;1,2}(z)$:
The matrix
$$
\mathbf{T}_{m;1,2}(z) =
 z^{m} \boldsymbol{\mu}_-^{(1)} + z^{m-1} \boldsymbol{\mu}_1 
 + \cdots + z \boldsymbol{\mu}_{m-1} + \boldsymbol{\mu}_+^{(2)}
$$
has the spectral curve \eqref{F-3m+1} with
$\deg f_1^\prime(z) = m-1, ~ \deg f_2^\prime(z) = 2m-1$
and $\deg f_3^\prime(z) = 3m-2$,
whose genus is $3m-2$.
The gauge matrix and the matrix \eqref{general-M} are
$$
  \mathbf{S} = 
  \begin{pmatrix}
    \vec{e}_1 \\
    \vec{e}_1 (\boldsymbol{\mu}_+^{(2)})^2 \\
    \vec{e}_1 \boldsymbol{\mu}_+^{(2)} 
  \end{pmatrix},
 ~~~
  \mathbf{M}(z) =
  z^m \begin{pmatrix}
       0 & 0 & 0 \\ 
       \ast & \ast & \ast \\
       \ast & \ast & \ast 
      \end{pmatrix}
   + \cdots +
  \begin{pmatrix}
       0 & 0 & 1 \\ 
       0 & 0 & 0 \\
       0 & 1 & 0 
      \end{pmatrix}.
$$

\noindent
(vi) $\mathbf{T}_{m;2,3}(z)$:
When $m \geq 3$, the matrix is defined as
$$
\mathbf{T}_{m;2,3}(z) =
 z^{m} \boldsymbol{\mu}_-^{(2)} + z^{m-1} \boldsymbol{\mu}_-^{(0)} 
 + z^{m-2} \boldsymbol{\mu}_2 + \cdots + z^2 \boldsymbol{\mu}_{m-2} 
 + z \boldsymbol{\mu}_+^{(0)} + \boldsymbol{\mu}_+^{(3)}. 
$$
Its spectral curve is given by \eqref{F-3m+2}
with $\deg f_1^\prime(z) = m-2, ~ \deg f_2^\prime(z) = 2m-3$
and $\deg f_3^\prime(z) = 3m-4$,
and the genus is $3m-4$.
The gauge matrix and the matrix \eqref{general-M} are obtained as 
\begin{equation}
  \label{3-vi}
  \mathbf{S} = 
  \begin{pmatrix}
    \vec{e}_3 \boldsymbol{\mu}_+^{(0)} \\
    \vec{e}_3 \boldsymbol{\mu}_+^{(0)} \boldsymbol{\mu}_-^{(2)} \\
    \vec{e}_3 
  \end{pmatrix},
  ~~
  \mathbf{M}(z) =
  z^{m} \begin{pmatrix}
       0 & 1 & 0\\ 
       0 & 0 & 0 \\
       0 & \ast & 0
      \end{pmatrix}
   + \cdots +
  z \begin{pmatrix}
       \ast & \ast & \ast \\
       \ast & \ast & \ast \\
       1 & 0 & 0 
      \end{pmatrix}
  + \begin{pmatrix}
     0 & 0 & 0 \\
     0 & 0 & \ast \\
     0 & 0 & 0
    \end{pmatrix}
\end{equation}
For the case of $m = 2$, we have
$$
\mathbf{T}_{2;2,3}(z) =
   z^2 \boldsymbol{\mu}_-^{(2)} 
  + z \,(\boldsymbol{\mu}_-^{(0)} \cap \boldsymbol{\mu}_+^{(0)}) 
 + \boldsymbol{\mu}_+^{(3)}, 
  ~~ \text{ where }
  (\boldsymbol{\mu}_-^{(0)} \cap \boldsymbol{\mu}_+^{(0)}) = 
  \begin{pmatrix}
        \ast & \ast & 0 \\
        \ast & \ast & \ast \\
        0 & \ast & \ast
      \end{pmatrix}.   
$$
Following this form, the gauge becomes
$$
  \mathbf{S} = \begin{pmatrix}
                 \vec{e_3}
                 (\boldsymbol{\mu}_-^{(0)} \cap \boldsymbol{\mu}_+^{(0)})  \\
                 \vec{e_3} 
                 (\boldsymbol{\mu}_-^{(0)} \cap \boldsymbol{\mu}_+^{(0)})  
                  \boldsymbol{\mu}_-^{(2)} \\
                 \vec{e_3}  
               \end{pmatrix}.
$$
The associated characteristic polynomial and the matrix \eqref{general-M}
are obtained by substituting $m=2$ in \eqref{F-3m+2} and \eqref{3-vi}. 

We construct the set of representatives 
$\{ \mathbf{M}(z) \}_F$ \eqref{M-rep}
based on $F(z,w)$ and $\mathbf{M}(z)$ for each cases.
One sees that in all cases 
the gauge matrices $\mathbf{S}$ reduce
the dimension of the variety $\{\mathbf{T}_{m;n_1,n_2}(z)\}_F$
by two.
The separation equation differs
depending on which of $\vec{e}_1$ and $\vec{e}_3$ 
the gauge matrix $\mathbf{S}$ includes.
For the cases of (i), (iii) and (v),
we have the invariant separation equation \eqref{separation} as 
$$
  B(z) = \Det 
  \begin{pmatrix}
    (T(z)_{1,2} , T(z)_{1,3}) \\
    (T(z)_{1,2},T(z)_{1,3})
    \begin{pmatrix}
      T(z)_{2,2} & T(z)_{2,3} \\
      T(z)_{3,2} & T(z)_{3,3}  
    \end{pmatrix}
  \end{pmatrix} = 0, 
  ~\text{where} ~ \vec{a} = (1,0,0),
$$ 
and for the rest cases,
$$
  B(z) = \Det 
  \begin{pmatrix}
    (T(z)_{3,1} , T(z)_{3,2}) \\
    (T(z)_{3,1},T(z)_{3,2})
    \begin{pmatrix}
      T(z)_{1,1} & T(z)_{1,2} \\
      T(z)_{2,1} & T(z)_{2,2}  
    \end{pmatrix}
  \end{pmatrix} = 0,
  ~\text{where} ~ \vec{a} = (0,0,1).
$$
In all cases $B(z)$ has a form as \eqref{separationB},
and each of them gives the eigenvalue by
\begin{align*}
  w_i = 
  \begin{cases}
   \Det 
   \begin{pmatrix}
      T(z_i)_{1,2} & T(z_i)_{1,3} \\
      T(z_i)_{3,2} & T(z_i)_{3,3}  
   \end{pmatrix}
   / T(z_i)_{1,2},  \text{ for (i),(iii),(v)},
  \\[5mm]
   \Det 
   \begin{pmatrix}
      T(z_i)_{1,1} & T(z_i)_{1,2} \\
      T(z_i)_{3,1} & T(z_i)_{3,2}  
   \end{pmatrix}
   / T(z_i)_{3,2},  \text{ for (ii),(iv),(vi)}.
  \end{cases}
\end{align*}
In conclusion, the separation equation uniquely determines
the effective divisor $P \in X(g) - D$
which is invariant under the gauge $\mathbf{S}$. 

%%%%%%%%%%%%%%%%%%%%%%%%%%%%%

\subsection{General $N$ cases}

In the case of general $N_{\geq 4}$,
we have $N(N-1)$ kinds of monodromy matrices $\mathbf{T}_{m;n_1,n_2}(z)$.
When $n_1 = n_2 = 1$,
the spectral curve $X$ is given by \eqref{curve-general}
where $\deg f_i(z) = im$, for $i=1,\cdots,N-1$, and $f_N(z) = Nm-1$.
Then the genus is $g = \frac{1}{2}(N-1)(Nm-2)$.
For each $\mathbf{T}_{m;1,1}(z)$ 
we have a gauge matrix \cite{Smirnov-Zeitlin0111};
\begin{align}
  \label{gauge-N}
  \mathbf{S} =
  \begin{pmatrix}
    \vec{e}_1 \\
    \vec{e_1} \boldsymbol{\mu}_1 (\boldsymbol{\mu}_-^{(1)})^{N-2} \\
    \vdots \\
    \vec{e_1} \boldsymbol{\mu}_1 \boldsymbol{\mu}_-^{(1)} \\
    \vec{e_1} \boldsymbol{\mu}_1
  \end{pmatrix},
\end{align}
which reduces the variety of $\{ \mathbf{T}_{m;1,1}(z)\}_F$ 
by $N-1$ dimensions.
Using the elements of $\mathbf{T}_{m;1,1}(z)$ given by
\begin{align*}
  \mathbf{T}_{m;1,1}(z) =
  \begin{pmatrix}
    a(z) & \vec{b}(z)\\
    \vec{c}(z)^T & \mathbf{d}(z)
  \end{pmatrix},
\end{align*}
the separation equation is defined as \cite{Scott94,Gekhtman95}
\begin{align*}
%  \label{B-eq}
  B(z)
  \equiv
  \Det
  \begin{pmatrix}
    \vec{b}(z) \\
    \vec{b}(z) \mathbf{d}(z) \\
    \vec{b}(z) \mathbf{d}(z)^2  \\
    \vdots \\
    \vec{b}(z) \mathbf{d}(z)^{N-2} \\
  \end{pmatrix}.
\end{align*}
Then $B(z)$ becomes a polynomial of $z$ of degree $g$,
and the zeros of $B(z)$ is invariant under the gauge transformation
induced by $\mathbf{S}$ \cite{Gekhtman95}.

Instead of showing other cases,
based on the above concrete studies
we introduce the conjecture for $\mathbf{S}$ as follows;
\begin{conjecture}
  \label{S-conjecture}
  For $\mathbf{T}_{m;n_1,n_2}(z)$ \eqref{T-general-form} there is  
  a gauge matrix $\mathbf{S}$ \eqref{M-rep} of the form
  \begin{align}
  \label{S-matrix}
  \mathbf{S} = 
    \begin{pmatrix}
       \vec{e}_1 \\
       \vec{e}_1 \boldsymbol{\mu}_a \boldsymbol{\mu}_b^{N-2} \\
         \vdots                         \\
       \vec{e}_1 \boldsymbol{\mu}_a \boldsymbol{\mu}_b \\
       \vec{e}_1 \boldsymbol{\mu}_a 
    \end{pmatrix}, \text{ for even } L,
     ~~~
    \begin{pmatrix}
       \vec{e}_N \boldsymbol{\mu}_a \\
       \vec{e}_N \boldsymbol{\mu}_a \boldsymbol{\mu}_b \\
         \vdots                         \\
       \vec{e}_N \boldsymbol{\mu}_a \boldsymbol{\mu}_b^{N-2} \\
       \vec{e}_N 
    \end{pmatrix}, \text{ for odd } L, 
  \end{align}
 where $\boldsymbol{\mu}_a, \boldsymbol{\mu}_b \in
  \{\boldsymbol{\mu}_-^{(n_1)}, \boldsymbol{\mu}_-^{(n_1-N+1)}, 
    \boldsymbol{\mu}_+^{(n_2)}, \boldsymbol{\mu}_+^{(n_2-N)}, 
  \boldsymbol{\mu}_j, ~j=2,\cdots,m-2\}$, 
such that $\mathbf{S}$ reduces  $\{\mathbf{T}_{m;n_1,n_2}(z)\}_F$
to a $g$-dimensional variety $\{\mathbf{M}(z)\}_F$ \eqref{M-rep}
and that
the associated separation equation \eqref{separation}
with the form as \eqref{separationB}
has the zeros invariant under $\mathbf{S}$.
\end{conjecture}
We briefly remark on the diagram \eqref{eigenvector-map}.
If we get the gauge matrix $\mathbf{S}$
which solves Problem \ref{problem}
then (a) becomes surjective,
since $\{ \mathbf{M}(z) \}_F$ is the set of representatives.
Therefore the map (c) exists such that the diagram  
\eqref{eigenvector-map} is commutative.

%%%%%%%%%%%%%%%%%%%%%%%%%%%%%%%%
%%%%%%%%%%%%%%%%%%%%%%%%%%%%%%%%

\section{Integrability of LV($N,L$)}

\subsection{Spectral curve and Poisson structure for LV($N,L$)}

We introduce the $N$ by $N$ Lax matrix for 
the extended Lotka-Volterra lattice \eqref{Bogo} as
\begin{equation}
  \label{Bogo-Lax}
  \Tilde{\mathbf{L}}_n(z)
    =
  \frac{1}{z  V_n^{\frac{N-1}{N}}}
  \Bigl( 
  \sum_{k=1}^{N-1} V_n \mathbf{E}_{k,k+1}
  + z^N (-1)^{N-1} \mathbf{E}_{N,1}
  + z^N (-1)^{N-2} \mathbf{E}_{N,2}
  \Bigr).
\end{equation}
We have much modified the original Lax matrix \cite{Bogo88},
and \eqref{Bogo-Lax} comes from $\overline{\mathbf{L}}_n(z)$ in 
\cite{Inoue02-JBogo}.
Note that $\Tilde{\mathbf{L}}_n(z)$ has been normalized as 
$\Det\Tilde{\mathbf{L}}_n(z) = 1$.
The monodromy matrix $\Tilde{\mathbf{T}}(z)$ of an $L$-periodic
model LV($N,L$) is defined as
\begin{equation}
  \label{monodromy-LV}
   \Tilde{\mathbf{T}}(z) =
    \prod_{k=1}^{\stackrel{L}{\curvearrowleft}}
                        \Tilde{\mathbf{L}}_k(z).
\end{equation}
The characteristic equation of $\Tilde{\mathbf{T}}(z)$,
\begin{equation}
  \label{curveX}
   \Det \bigl( w \openone - \Tilde{\mathbf{T}}(z) \bigr) = 0,
\end{equation}
gives an algebraic curve $\Tilde{X}$.
For this equation we have the automorphism $\tau$ of order $N$,
$$
 \tau : ~ (z,w) \mapsto (\epsilon z ,\epsilon^{-k_2} w),
$$
where $\epsilon = \mathrm{e}^{\frac{2 \pi \mathrm{i}}{N}}$, and 
$k_2$ is defined by \eqref{k-s}.
We define the matrix $\mathbf{T}_{LV}(z)$,
\begin{align}
  \label{T-LV}
\mathbf{T}_{LV}(z) \equiv z^{\frac{k_2}{N}} 
                          \Tilde{\mathbf{T}}(z^\frac{1}{N}),
\end{align}
then its matrix elements become polynomials of $z$ 
and $\Det \mathbf{T}_{LV}(z) = z^{k_2}$.
The characteristic equation of $\mathbf{T}_{LV}(z)$
gives the quotient curve $\Tilde{X}/\tau$.

On the other hand, 
the Hamiltonian structure of LV($N,L$) is defined by the Poisson brackets
\cite{Bogo88}
\begin{equation} 
  \label{PoissonLV}
  \{ V_n, V_m \}
  =
  2 \,V_n V_m \sum_{k=1}^{N-1} ( \delta_{m,n+k} - \delta_{m,n-k} ),
\end{equation}
and the Hamiltonian $H_1 = \sum_{n=1}^L V_n$.
Using these settings,
the time evolution \eqref{Bogo} is given by 
$$
   \frac{\partial V_n}{\partial t_1}
  =
  \{ V_n ~,~ H_1 \}
$$    
with $t = t_1$.
We let $\mathcal{A}_{LV}$ be the Poisson bracket algebra for
$\mathbb{C}[V_n; n \in \mathbb{Z}/L\mathbb{Z}]$
whose defining relations are given by \eqref{PoissonLV}.
We have the center of $\mathcal{A}_{LV}$ 
denoted by $\mathcal{A}_{LV}^0$ as follows; 
\begin{proposition}
  \label{LV-center}
  The center $\mathcal{A}_{LV}^0$ is generated by the variables
  \begin{align}
  \mathcal{P}_{k}^{(i)} = \prod_{n=0}^{\frac{L}{k}-1} (V_{k n+i}),
   ~~ \text{for } k \in \mathcal{K},~ i \in \{1,\cdots, k\},
  \end{align}
where
\begin{align}
  \label{k-K}
  \mathcal{K} = \{ k \in \{ 1, \cdots, N \} 
                   ~|~~ k|N \text{ or } k|(N-1) \}
                \sqcap \{ k ~|~~ k|L \}.
\end{align}  
Here $ k | L$ means that $L$ is a multiple of $k$.
\end{proposition}
See Appendix B for the proof.
Since the set $\{ \mathcal{P}_{k}^{(i)} | i \in \{1,\cdots, k\} \}$
is generated by $\{ \mathcal{P}_{k^\prime}^{(j)} | 
j \in \{1, \cdots, k^\prime\} \}$
for $k | k^\prime$,
to generate $\mathcal{A}_{LV}^0$ it is enough to have a set
$$
  \{ \mathcal{P}_{k}^{(i)} ~|~
       k \in \mathcal{K}_0, ~i \in \{1,\cdots, k\} \},
$$
where
$
 \mathcal{K}_0 = \{ \text{max} [ k \in \mathcal{K} \text{ for } k | N ], ~
                    \text{max} [k \in \mathcal{K} \text{ for } k | (N-1)] \}.
$
Then the number of independent generators of $\mathcal{A}_{LV}^0$ is 
\begin{equation}
  \label{number-center}
  n_0 = \sum_{k \in \mathcal{K}_0} k - ( | \mathcal{K}_0 | - 1).
\end{equation}
Based on the structure of the monodromy matrix \eqref{monodromy-LV},
we introduce a variable  
$$
  \mathcal{P}_0 \equiv \prod_{n=1}^L (V_n)^{- \frac{1}{N}} 
    = \bigl(\mathcal{P}_1^{(1)}\bigr)^{-\frac{1}{N}},
$$
which is Poisson commutative with any $V_n$.
Therefore $\mathcal{A}_{LV}$ is naturally extended to the
Poisson bracket algebra over 
$\mathbb{C}(\mathcal{P}_0,V_n ; n \in \mathbb{Z}/L\mathbb{Z})$.
We denote this algebra by $\mathcal{A}_{LV}^\prime$.

A family of the integrals of motion(IM) for LV($N,L$)
which includes the Hamiltonian $H_1$ appears as coefficients of 
the characteristic equation \eqref{curveX}.
\begin{proposition} \cite{InoueHikami98-Bogo,HikamiInoueKomori99}
  \label{IM-commute}
  The IM compose the commuting subalgebra of $\mathcal{A}^\prime_{LV}$.  
\end{proposition}
{\em Proof}.
We show the outline of the proof.
We introduce the variable transformation
\begin{equation*}
%  \label{V-trans}
  V_n = (P_n P_{n+1} \cdots P_{n+N-1})^{-1} Q_n^{-1} Q_{n+N-1},
\end{equation*}
where $P_n$, $Q_n$ are canonical variables,
\begin{align}
  \label{canonicalPQ}
  \{ P_n ~,~ Q_m \}
  = \delta_{n,m} P_n \, Q_n  ,
  ~~~~
  \{ P_n ~,~ P_m \}
  =
  \{ Q_n ~,~ Q_m \} = 0.
\end{align}
Then the matrix $\mathbf{T}_{LV}(z)$ is transformed as
\begin{align}
  \label{gaugeB}
  \mathbf{T}_C(z) = \mathbf{B}_1 \mathbf{T}_{LV}(z) (\mathbf{B}_1)^{-1},
\end{align}
using a diagonal matrix 
$\mathbf{B}_1 = \mathbf{B}_1(P_1,\cdots,P_{N-1},Q_1,\cdots,Q_{N-1})$. 
The matrix $\mathbf{T}_C(z)$ turns out to satisfy the 
fundamental Poisson relation \eqref{T-Poisson}
\begin{align}
  \label{TC-Poisson}
  \{\mathbf{T}_C(z)
  \stackrel{\otimes}{,}
  \mathbf{T}_C(z^\prime) \}
  =
  [\, \mathbf{r}(z/z^{\prime}) ~,~ \mathbf{T}_C(z) \otimes
  \mathbf{T}_C(z^\prime) \,],
\end{align}
with the $r$-matrix \eqref{classical-r}.
See  \cite{HikamiInoueKomori99,Inoue02-JBogo} for details of 
the gauge matrix $\mathbf{B}_1$ and $\mathbf{T}_C(z)$.
Note that the characteristic equation for the matrix $\mathbf{T}_C(z)$
is obtained from \eqref{curveX} by a transformation 
$ 
%X \to X_{\tau} :~ 
(z,w) \mapsto (z^{\frac{1}{N}},w z^{\frac{k_2}{N}}),
$
and that the coefficients of the characteristic polynomial belong to
$\mathbb{C}[\mathcal{P}_0,V_n; n \in \mathbb{Z}]$.
Then the proposition follows. $\square$

We introduce 
a grading on $\mathcal{A}_{LV}$ as $\deg V_n = 1$.
Since the IM are obtained as homogeneous polynomials of $V_n$,
we can identify each of IM based on the grading.
For instance, for the Hamiltonian $H_1$ we have  $\deg H_1 = 1$.
Let $n_H$ be the number of the independent elements of IM
in $\mathcal{A}_{LV}$. 
By putting the IM in the order of the grading,
we obtain 
\begin{equation}
  \label{IM}
  H_1 , H_2 , \cdots, H_{n_H}.
\end{equation}
The Proposition \ref{IM-commute} yields 
\begin{corollary}
  The family of IM generate $n_H$ independent flows
  for LV($N,L$) defined as
  \begin{equation}
  \label{time-evol}
  \frac{\partial \mathcal {O}}{\partial t_i}
  \equiv \{  \mathcal{O} ~,~ H_i \},
  ~~~ \text{for~ } \mathcal{O} \in \mathcal{A}_{LV}^\prime, 
     ~i = 1, \cdots, n_H.
  \end{equation}
\end{corollary}

We comment that in \cite{Suris94} the Hamiltonian structure of LV($N,L$)
is studied by applying the $r$-matrix method \cite{ReySemenov94} to
the {\em big} Lax matrix of $L$ by $L$, 
and the involution of IM is clarified by this approach.  
Since our aim here is to establish the eigenvector map for LV($N,L$) 
based on the monodromy matrix \eqref{T-Poisson}, 
it is important to get the {\em small} monodromy matrix of $N$ by $N$ with 
the fundamental Poisson relation \eqref{TC-Poisson}.

%%%%%%%%%%%%%%%

\subsection{Realization of $\mathbf{M}_F(z)$
and the integrable structure of LV($N,L$)}

We find that the matrix $\mathbf{T}_C(z)$ \eqref{gaugeB} gives
the realization of $\mathbf{T}_{m;n_1,n_2}(z)$ \eqref{T-general-form},
namely both of the form and the Poisson structure of $\mathbf{T}_C(z)$
coincide with those of $\mathbf{T}_{m;n_1,n_2}(z)$.
We obtain a similar relation as \eqref{LT-corresp} as follows;
\begin{proposition}
Under a condition 
\begin{align}
  \label{det-condition}
  \Det\mathbf{T}_{m;n_1,n_2}(z) =z^{n_2-1}, 
\end{align}
$\mathbf{T}_C(z)$ realizes $\mathbf{T}_{m;n_1,n_2}(z)$ and they are related as
\begin{align}
  \label{TC-Tm}
  \mathbf{T}_C(z) 
    = 
    \begin{cases}
      \mathbf{T}_{m;1,1}(z), ~\text{ for } k_1 = k_2 = 0,
      \\ 
      \mathbf{T}_{m+1;N-k_1,k_2+1}(z), 
      ~\text{ for } k_1, k_2 \neq 0, ~0 \leq k_1 - k_2 \leq N-2,
      \\
      \mathbf{T}_{m+2;N-k_1,k_2+1}(z), ~\text{ for } k_1 - k_2 \leq -1.
    \end{cases}    
\end{align}
\end{proposition}
{\em Proof.}
First we check the coincidence of the form.
Note that the condition \eqref{det-condition} comes from the
normalization of $\Tilde{\mathbf{L}}_n(z)$. 
The Lax matrices 
$z^{\frac{1}{N}} \Tilde{\mathbf{L}}_n(z^{\frac{1}{N}})$
\eqref{Bogo-Lax} and $\mathbf{L}_n(z)$ \eqref{general-Lax} have 
the same form as polynomial matrices.
Then we see that
$\mathbf{T}_{LV}(z)$ \eqref{T-LV} and 
$z^{-\frac{L}{N} + \frac{k_2}{N} + m_2} \mathbf{T}^{(L)}(z)$ \eqref{T-L}
has a same form.
Since the gauge $\mathbf{B}_1$ \eqref{gaugeB} is diagonal and does not change 
the form of $\mathbf{T}_{LV}(z)$,
we obtain the correspondence of  
$\mathbf{T}_C(z) = \mathbf{B}_1 \mathbf{T}_{LV}(z) \mathbf{B}_1^{-1}$ 
and $\mathbf{T}^{(L)}(z)$.
By using Lemma \ref{T-Lproduct} and the relation $L=N m_2 + k_2$
\eqref{k-s},
finally we obtain \eqref{TC-Tm}.

Next, we observe the Poisson structure.
The condition \eqref{det-condition}
and \eqref{T-Poisson} do not contradict each other,
since Proposition \ref{prop:Poisson} says that 
$\Det\mathbf{T}_{m;n_1,n_2}(z)$ belongs to the center of $\mathcal{A}_C$.
Then from \eqref{T-Poisson} and \eqref{TC-Poisson},
the monodromy matrices $\mathbf{T}_C(z)$ 
obviously has the same Poisson structure as that of 
$\mathbf{T}_{m;n_1,n_2}(z)$. 
$\square$
\\[2mm]
This Proposition is the reason why we denoted the Poisson bracket algebra of
$\mathbf{T}_{m;n_1,n_2}(z)$ using $\mathcal{A}_C$ in \S 2. 
Once we associate $\mathbf{T}_C(z)$ to 
$\mathbf{T}_{m;n_1,n_2}(z)$,
$\mathbf{T}_C(z)$ realizes $\{\mathbf{T}_{m;n_1,n_2}(z)\}_F$
where $\mathbf{T}_{m;n_1,n_2}^0(z)$ corresponds to
the initial condition for $\mathbf{T}_C(z)$. 
We also see $\Tilde{X} / \tau \simeq X$.

In the following we discuss the integrability of LV($N,L$)
based on the representative $\{\mathbf{M}(z)\}_F$ 
\eqref{M-rep}
and the Poisson bracket algebra $\mathcal{A}_{M}$ 
realized by LV($N,L$). We introduce an important proposition;
\begin{proposition}
  \label{prop:subset}
  If the gauge matrix $\mathbf{S}$ which meets the conditions in
  Conjecture \ref{S-conjecture} exists,
  then 
  (1) $\mathcal{A}_{M} \subset \mathcal{A}_{LV}^\prime$,
  (2) the separation equation \eqref{separation} gives 
      $g$ algebraic relations 
      between $z_i$ $(i = 1, \cdots, g)$ and 
      $V_n$ $(n \in \mathbb{Z}/L \mathbb{Z})$.
\end{proposition}
Remember that the matrix $\mathbf{T}_C(z)$ is no longer written in terms of
the dynamical variables of LV($N,L$),
but of the canonical variables \eqref{canonicalPQ}.
Therefore $\mathcal{A}_{M} \subset \mathcal{A}_C$ is trivial
but the claim (1) in the above proposition is not.  
This claim was conjectured in \cite{Inoue02-JBogo} 
and now is proved in a simple way. 
We add the proof of Proposition \ref{prop:subset} at Appendix C.

On the tangent space of $\boldsymbol{\mathcal{M}}_F$ 
there is the $g$ dimensional invariant vector field
which induces
the evolution of the divisor $P$ \eqref{div-P}
linearized on $J_{\text{aff}}(X)$.
When $n_H$ is equal to $g$,
we can identify the coordinates on  $J_{\text{aff}}(X)$ with
the times $t_i$ \eqref{time-evol}, and 
get $z_i$ as a functions of $t_i$;
$z_i = z_i(t_1, \cdots t_g)$. 
Further, if $n_H = \frac{1}{2}(L - n_0)$ is satisfied,
we can reduce the integrability of LV($N,L$) to 
$L$ independent algebraic relations between 
the dynamical variables of LV($N,L$) and
$H_i$ \eqref{IM}, $z_i$ \eqref{separation} 
and $n_0$ generators of $\mathcal{A}_{LV}^0$ \eqref{number-center}.
We summarize the integrability of LV($N,L$) as follows;
\begin{proposition}
  \label{def:integrability}
      LV($N,L$) is algebraic completely integrable 
     if 
     \begin{align}
       \label{IM-genus}
     g = n_H = \frac{1}{2}(L - n_0) 
     \end{align} 
     and  
     Proposition \ref{prop:subset} is satisfied.    
\end{proposition}

In \S 3, we solved Problem \ref{problem} for the cases
of $N=2,3$ and the special case of general $N$.
We obtained 
the gauge matrices $\mathbf{S}$ \eqref{M-rep}
which meet Conjecture \ref{S-conjecture},
then Proposition \ref{prop:subset} is satisfied for these cases.
The last case corresponds to
LV($N,L$) with the special periodicity $L = N(N-1)m$
studied in ref. \cite{Inoue02-JBogo} 
where \eqref{IM-genus} was proved and 
Proposition \ref{prop:subset} was supposed.
Now we have Proposition \ref{prop:subset} satisfied, then
we conclude that 
\begin{theorem}
LV($N,N(N-1)m$) is algebraic completely integrable.
\end{theorem}
In the following, we investigate Propositions \ref{prop:subset} 
and \ref{def:integrability}
for the results in \S 3 and show Theorem \ref{th:N=2-3}.

%%%%%%%%%%%%%%%%%

\subsection{LV($2,L$)}

Depending on the periodicity $L$ we have two cases;

\noindent
(i) $L=2m$, $\mathbf{T}_C(z) = \mathbf{T}_{m;1,1}(z)$: 
The IM are obtained as the coefficients of \eqref{ch-2-even} with
$$
  f_1(z) = \mathcal{P}_0 \bigl(z^m + H_1 z^{m-1} + H_2 z^{m-2} - \cdots 
                 + z H_{m-1} 
                 + (\mathcal{P}_2^{(1)} + \mathcal{P}_2^{(2)}) \bigr).
$$ 
Here we have $m-1$ independent IM identified by their degree, 
$\deg H_i = i$.
The center $\mathcal{A}_{LV}^0$ is generated by two 
of $\mathcal{P}_1^{(1)}, \mathcal{P}_2^{(1)}$ and $\mathcal{P}_2^{(2)}$.
The genus of $X$ is equal to $n_H$.

\noindent
(ii) $L=2m+1$,  $\mathbf{T}_C(z) = \mathbf{T}_{m+1;1,2}(z)$: 
We have $m$ independent IM given by \eqref{ch-2-odd} with
$$
  f_1^\prime(z) = \mathcal{P}_0 \bigl( z^m - H_1 z^{m-1} + H_2 z^{m-2} - \cdots 
                 + (-)^{m} H_{m} \bigr), 
$$
where $\deg H_i = i$.
The center $\mathcal{A}_{LV}^0$ is generated by 
$\mathcal{P}_1^{(1)}$ only.

In both cases \eqref{IM-genus} is satisfied 
and the gauge matrices $\mathbf{S}$ \eqref{gauge-2-even} and \eqref{gauge-2-odd}
fulfill
Proposition \ref{prop:subset}. 
Therefore we conclude that LV($2,L$) is algebraic completely integrable.
The correspondence of the periodicity $L$ and the genus $g$ is 
summarized as 
\begin{center}
\begin{tabular}{ |c|l l l l l l l c c l | }
  \hline  
  $L$ & $3$ & $4$ & $5$ & $6$ & $7$ & $8$ & $\cdots$ & $2m$ & $2m+1$ 
  & $\cdots$ \\
  \hline
  $g$ & $1$ & $1$ & $2$ & $2$ & $3$ & $3$ & $\cdots$ & $m-1$ & $m$ 
  & $\cdots$ \\
  \hline
\end{tabular} ~ .
\end{center}

%%%%%%%%%%%%%%%%%%%%%%%%%%%%%%%%%%

\subsection{LV($3,L$)}

The periodicity $L$ is classified into $6$ cases;

\noindent
(i) $L=6m$, $\mathbf{T}_C(z) = \mathbf{T}_{m;1,1}(z)$:
The IM are obtained as
\begin{align*}
  &f_1(z) = \mathcal{P}_0^2 ( f_{3m} z^m +  f_{3m+1} z^{m-1} + \cdots 
                             + f_{4m} ), 
  \\
  &f_2(z) = \mathcal{P}_0 ( z^{2m} +  f_1 z^{2m-1} + \cdots + f_{2m} ),
\end{align*}  
where we set $f_i$ so as to accomplish $\deg f_i = i$. 
The generators of $\mathcal{A}_{LV}^0$ have the ordering as
$\deg \mathcal{P}_2^{(i)} = 3m, ~ \deg \mathcal{P}_3^{(i)} = 2m$,
then 
$f_{3m}$, $f_{4m}$ and $f_{2m}$ 
belong to $\mathcal{A}_{LV}^0$.
Actually, we have relations
\begin{align}
  \label{f-P-2}
  &z^2 + f_{3m} z + \mathcal{P}_1^{(1)} 
  = (z - \mathcal{P}_2^{(1)}) (z - \mathcal{P}_2^{(2)}),
  \\
  \label{f-P-3}
  &z^3 + z^2 f_{2m} + z f_{4m} + \mathcal{P}_1^{(1)} 
  = (z - \mathcal{P}_3^{(1)}) (z - \mathcal{P}_3^{(2)})
    (z - \mathcal{P}_3^{(1)}).
\end{align}
In conclusion we have $n_H = 3m-2$ which is equal to $g$,
and $n_0 = 4$.   
   
\noindent
(ii) $L=6m+1$, $\mathbf{T}_C(z) = \mathbf{T}_{m+1;2,2}(z)$: 
We have
\begin{align*}
  &f_1^\prime(z) = 
  \mathcal{P}_0^2 ( f_{3m+1} z^{m-1}  + f_{3m+2} z^{m-2}  + \cdots 
                             + f_{4m} ), 
  \\
  &f_2^\prime(z) = \mathcal{P}_0 ( z^{2m} + f_1 z^{2m-1}  + \cdots + f_{2m} ).
\end{align*}  
In this case we have only a generator of $\mathcal{A}_{LV}^0$;
$\mathcal{P}_1^{(1)}$,
and no $f_i$ belongs to $\mathcal{A}_{LV}^0$.
Then $n_H = 3m$ and $n_0 = 1$. 

\noindent
(iii) $L=6m+2$, $\mathbf{T}_C(z) = \mathbf{T}_{m+1;1,3}(z)$: 
\begin{align*}
  &f_1^\prime(z) = \mathcal{P}_0^2 ( f_{3m+1} z^{m-1}  + f_{3m+2} z^{m-2}  + \cdots 
                             + f_{4m+1} ), 
  \\
  &f_2^\prime(z) = \mathcal{P}_0 ( z^{2m} + f_1 z^{2m-1}  + \cdots + f_{2m} ).
\end{align*}  
Since $\deg \mathcal{P}_2^{(i)} = 3m+1$,
we see $f_{3m+1} \in \mathcal{A}_{LV}^0$, 
which satisfies a relation similar to \eqref{f-P-2}.
Then we have $n_H = 3m$ and $n_0 = 2$. 

\noindent
(vi) $L=6m+3$, $\mathbf{T}_C(z) = \mathbf{T}_{m+1;2,1}(z)$: 
\begin{align*}
  &f_1(z) = \mathcal{P}_0^2 ( f_{3m+2} z^{m}  + f_{3m+2} z^{m-1}  + \cdots 
                             + f_{4m+2} ), 
  \\
  &f_2(z) = \mathcal{P}_0 ( z^{2m+1} + f_1 z^{2m}  + \cdots + f_{2m+1} ).
\end{align*}  
Since $\deg \mathcal{P}_2^{(i)} = 2m+1$,
we see $f_{4m+2}, f_{2m+1} \in \mathcal{A}_{LV}^0$, 
which satisfy a relation similar to \eqref{f-P-3}.
Then we have $n_H = 3m$ and $n_0 = 3$. 

\noindent
The remaining cases,
\\
(v)
$L=6m+4$, $\mathbf{T}_C(z) = \mathbf{T}_{m+1;1,2}(z)$
\\
(iv)
$L=6m+5$, $\mathbf{T}_C(z) = \mathbf{T}_{m+2;2,3}(z)$
\\
permit the same analysis.

For all $L$ 
we have $n_H$ and $n_0$ which satisfy \eqref{IM-genus}. 
Remember that in \S 3.2 we have constructed the 
$\{\mathbf{M}(z)\}_F$ with the gauge matrices $\mathbf{S}$
which meet Proposition \ref{prop:subset}.
Herewith we prove the algebraic completely integrability
of LV($3,L$).
As same as the $N=2$ case,
we summarize the correspondence of $L$ and $g$;
\begin{center}
\begin{tabular}{ |l|c c c c c c c c c c c c c| }
  \hline  
  $L$ & $5$ & $6$ & $7$ & $8$ & $9$ & $10$ & $\cdots$ &
  $6m$ & $6m+1$ & $6m+2$ & $6m+3$ & $6m+4$ & $6m+5$ \\
  \hline
  $g$ & $2$ & $1$ & $3$ & $3$ & $3$ & $4$ & $\cdots$ &
  $3m-2$ & $3m$ & $3m$ & $3m$ & $3m+1$ & $3m+2$ \\
  \hline
\end{tabular} ~.
\end{center}

%%%%%%%%%%%%%%%%%%%%%%%%%%%%%%%%%%%%
%%%%%%%%%%%%%%%%%%%%%%%%%%%%%%%%%%%%

\subsection*{Acknowledgements}

The author thanks Prof. A. Nakayashiki for informing about 
\cite{Vanhae01,Van1638}.
She appreciates discussion with T. Takenawa and T. Yamazaki.
She also thanks the referees for valuable comments
which have much improved the manuscript.
R. I. is a Research Fellow of the Japan Society for the Promotion of Science.

%%%%%%%%%%%%%%%%%%%%%%%%%%%%%%%%%%%%%
%%%%%%%%%%%%%%%%%%%%%%%%%%%%%%%%%%%%%

\newpage

\setcounter{equation}{0}

\renewcommand{\theequation}{A.\arabic{equation}}

\addcontentsline{toc}{chapter}{Appendix \protect\numberline{A}
                               Proof of Lemma \ref{T-Lproduct}}

\subsection*{Appendix A ~ Proof of Lemma \ref{T-Lproduct}}

We show the outline of the proof.
We use the integers defined at \eqref{m-s} and \eqref{k-s},
and set a matrix $\mathbf{T}^{(L)}(z)$;
\begin{align}
  \label{T-L}
  \mathbf{T}^{(L)}(z) = z^{-m_2} \prod_{n=1}^{L} \mathbf{L}_n(z).
\end{align}
By definition, first we have 
$$
  \mathbf{T}^{(1)}(z) 
  = \mathbf{L}_1(z)
  = \boldsymbol{\mu}_-^{(N-1)} z
  + (\boldsymbol{\mu}_-^{(0)} \cap \boldsymbol{\mu}_+^{(2)}).
$$
Therefore we obtain the correspondence 
$\mathbf{T}^{(1)}(z) = \mathbf{T}_{1;N-1,2}$. 
Assume $\mathbf{T}^{(L)}(z) = \mathbf{T}_{m;n_1,n_2}(z)$. 
When we set $\mathbf{T}^{(L)}(z) = (t^{(L)}_{i,j})_{1\leq i,j\leq N}$,
$\mathbf{T}^{(L+1)}(z)$ are related to $\mathbf{T}^{(L)}(z)$ as
\begin{align*}
  \mathbf{T}^{(L+1)}(z) 
  = 
  \begin{cases}
  &\displaystyle{\sum_{j=1}^{N}}  
       \Bigl(\displaystyle{\sum_{i=1}^{N-1}} 
     \mathbf{E}_{i,j} l_i^{(L+1)} t^{(L)}_{i+1,j} 
      + z \mathbf{E}_{N,j} 
          (l_N^{(L+1)} t^{(L)}_{1,j} + l_0^{(L+1)} t^{(L)}_{2,j}) 
       \Bigr), \text{ for } n_2 \neq N,
  \\
  &\displaystyle{\sum_{j=1}^{N}}  
    \Bigl( \frac{1}{z} \displaystyle{\sum_{i=1}^{N-1}} 
     \mathbf{E}_{i,j} l_i^{(L+1)} t^{(L)}_{i+1,j} 
      + \mathbf{E}_{N,j} 
          (l_N^{(L+1)} t^{(L)}_{1,j} + l_0^{(L+1)} t^{(L)}_{2,j}) 
       \Bigr), \text{ for } n_2 = N,
  \end{cases}
\end{align*}
then we find the correspondence
\begin{align*}
  \mathbf{T}^{(L+1)}(z) 
   =
   \begin{cases}
      \mathbf{T}_{m;n_1-1,n_2+1}(z),  \text{ for } n_1 \neq 1, n_2 \neq N,
      \\
      \mathbf{T}_{m+1;N-1,n_2+1}(z),  \text{ for } n_1 = 1, n_2 \neq N,
      \\
      \mathbf{T}_{m;N-1,1}(z),  \text{ for } n_1 = 1, n_2 = N,
      \\
      \mathbf{T}_{m-1;n_1-1,1}(z),  \text{ for } n_1 \neq 1, n_2 = N.
   \end{cases}
\end{align*}
By induction, we obtain \eqref{LT-corresp}. ~~ $\square$

\setcounter{equation}{0}

\renewcommand{\theequation}{B.\arabic{equation}}

\addcontentsline{toc}{chapter}{Appendix \protect\numberline{B}
                               Proof of Proposition \ref{LV-center}}

%%%%%%%%%%%%%%%%%%%%%%%%%%%%%

\subsection*{Appendix B ~ Proof of Proposition \ref{LV-center}}

Based on the periodicity $L$ and the Poisson relations \eqref{PoissonLV},
we can set candidates for the generators of $\mathcal{A}_{LV}^0$ as
$$
  \mathcal{P}_{k}^{(i)} = \prod_{n=0}^{\frac{L}{k}-1} (V_{k n+i}),
   ~~ \text{for } k \in \{1, \cdots, N\},~ k | L \text{ and } 
  i \in \{1,\cdots,k\}. 
$$
Our goal is to determine $k$.
The condition for
a variable $\mathcal{P}_{k}^{(i)}$ to belong to $\mathcal{A}_{LV}^0$;
$$
   \{ V_n ~,~ \mathcal{P}_{k}^{(i)} \} = 0, 
   ~~\text{ for } n \in \mathbb{Z}/L \mathbb{Z},
$$
reduces to
\begin{align}
  \label{condition-0}
   \sum_{m \in \mathbb{Z}/L\mathbb{Z}, ~m = i \text{ mod } k} ~ 
   \sum_{l=1}^{N-1} 
   ( \delta_{m,n+l} - \delta_{m,n-l}) = 0.
\end{align}   
Assume that 
we have $2j$ non-zero terms in the summation of \eqref{condition-0}
for $j \in \{1, \cdots, N-1\}$,
where 
$j$ of them offer $+1$ and the others offer $-1$.
In the case of $j=1$ we easily obtain 
$k=N$ if $N|L$ is satisfied, and $k=N-1$ if $(N-1)|L$.
In the case of $j=N-1$ we have $k=1$ for all $L$. 
In the following, we study the cases of $2 \leq k \leq N-2$.

Without limiting the generality, we consider the $n=0$ case in 
\eqref{condition-0}.
Let $m = n_0$ in \eqref{condition-0} be 
the leftmost lattice point where the first $-1$ occurs
for $-(N-1) \leq n_0 \leq -N+k$.
In $j=2$ case,
the condition for $k$ \eqref{condition-0} is reduced to
\begin{equation}
  \label{condition-1}
n_0 + k < 0 ~ \text{ and } ~ N-k \leq n_0 + 3k \leq N-1.
\end{equation}
This situation is depicted as
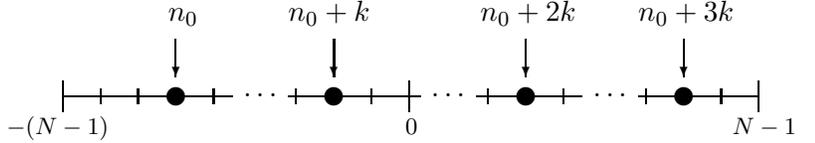
\begin{center}
  \unitlength=5mm
  \begin{picture}(20,3)(0,-0.5)
  \put(0,0){\line(1,0){4.5}}
  \put(4.8,0){$\dots$}
  \put(6,0){\line(1,0){3.5}}
  \put(9.8,0){$\dots$}
  \put(11,0){\line(1,0){2.8}}
  \put(14.1,0){$\dots$}
  \put(15.3,0){\line(1,0){3.2}}
  \put(0,-0.4){\line(0,1){0.8}}
  \put(-1.5,-1){\footnotesize{$-(N-1)$}}
  \put(1,-0.2){\line(0,1){0.4}}
  \put(2,-0.2){\line(0,1){0.4}}
  \put(3,0){\circle*{0.5}}
  \put(3,1.5){\vector(0,-1){1}}
  \put(2.8,2){$n_0$}
  \put(4,-0.2){\line(0,1){0.4}}
  \put(6.2,-0.2){\line(0,1){0.4}}
  \put(7.2,0){\circle*{0.5}}
  \put(7.2,1.5){\vector(0,-1){1}}
  \put(6,2){$n_0+k$}
  \put(8.2,-0.2){\line(0,1){0.4}}
  \put(9.2,-0.4){\line(0,1){0.8}}
  \put(9.1,-1){\footnotesize{$0$}}
  \put(9.2,-0.2){\line(0,1){0.4}}
  \put(11.3,-0.2){\line(0,1){0.4}}
  \put(12.3,0){\circle*{0.5}}
  \put(12.3,1.5){\vector(0,-1){1}}
  \put(11.1,2){$n_0+2k$}
  \put(13.3,-0.2){\line(0,1){0.4}}
  \put(15.5,-0.2){\line(0,1){0.4}}
  \put(16.5,0){\circle*{0.5}}
  \put(16.5,1.5){\vector(0,-1){1}}
  \put(15.3,2){$n_0+3k$}
  \put(16.5,-0.2){\line(0,1){0.4}}
  \put(17.5,-0.2){\line(0,1){0.4}}
  \put(18.5,-0.4){\line(0,1){0.8}}
  \put(17.8,-1){\footnotesize{$N-1$}}
  \end{picture}.
\end{center}
Here black circles mean where the non-zero terms are offered in
\eqref{condition-0}.
We have two critical cases for $n_0$;
\\
(i) when $n_0 = - (N-1)$, \eqref{condition-1} reduces to 
\begin{equation}
  \label{1-1}
  \frac{2N-1}{4} \leq k \leq \frac{2(N-1)}{3}. 
\end{equation}
(ii) When $n_0 = - N+k$, \eqref{condition-1} becomes
\begin{equation}
  \label{1-2}
  \frac{2N}{5} \leq k \leq \frac{2N-1}{4}.
\end{equation}
Since $\frac{2N-1}{4} \not\in \mathbb{Z}$, \eqref{1-1} and \eqref{1-2} are not
satisfied at the same time.
When $k$ satisfies (i), we should relate this $k$ to a condition
\\ 
(i') when $n_0 = - N+k$,
$n_0 + k = 0$ is imposed;
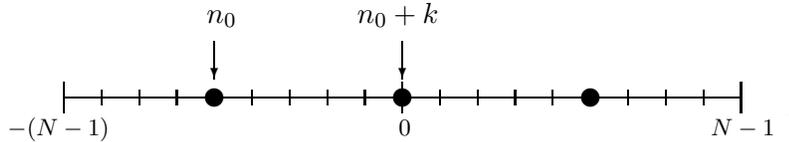
\begin{center}
  \unitlength=5mm
  \begin{picture}(19,3)(0,-0.5)
  \put(0,0){\line(1,0){18}}
  \put(0,-0.4){\line(0,1){0.8}}
  \put(-1.5,-1){\footnotesize{$-(N-1)$}}
  \put(1,-0.2){\line(0,1){0.4}}
  \put(2,-0.2){\line(0,1){0.4}}
  \put(3,-0.2){\line(0,1){0.4}}
  \put(4,0){\circle*{0.5}}
  \put(4,1.5){\vector(0,-1){1}}
  \put(3.8,2){$n_0$}
  \put(4,-0.2){\line(0,1){0.4}}
  \put(5,-0.2){\line(0,1){0.4}}
  \put(6,-0.2){\line(0,1){0.4}}
  \put(7,-0.2){\line(0,1){0.4}}
  \put(8,-0.2){\line(0,1){0.4}}
  \put(9,-0.4){\line(0,1){0.8}}
  \put(8.9,-1){\footnotesize{$0$}}
  \put(9,0){\circle*{0.5}}
  \put(9,1.5){\vector(0,-1){1}}
  \put(7.8,2){$n_0+k$}
  \put(9,-0.2){\line(0,1){0.4}}
  \put(10,-0.2){\line(0,1){0.4}}
  \put(11,-0.2){\line(0,1){0.4}}
  \put(12,-0.2){\line(0,1){0.4}}
  \put(13,-0.2){\line(0,1){0.4}}
  \put(14,-0.2){\line(0,1){0.4}}
  \put(14,0){\circle*{0.5}}
  \put(15,-0.2){\line(0,1){0.4}}
  \put(16,-0.2){\line(0,1){0.4}}
  \put(17,-0.2){\line(0,1){0.4}}
  \put(18,-0.4){\line(0,1){0.8}}
  \put(17.2,-1){\footnotesize{$N-1$}}
  \end{picture} .
\end{center}
Then we obtain $k = \frac{N}{2}$,
which turns out to be the $j=1$ case. 
\\[2mm]
On the other hand, 
when $k$ satisfies (ii), we relate it to
\\ 
(ii') when $n_0 = -(N-1)$,  $n_0 + 2k = 0$ is required;
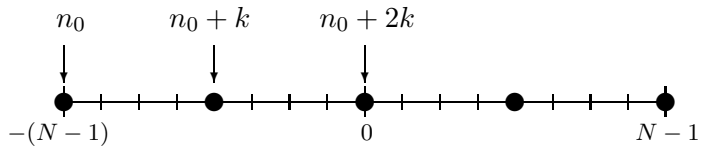
\begin{center}
  \unitlength=5mm
  \begin{picture}(17.2,3)(0,-0.5)
  \put(0,0){\line(1,0){16}}
  \put(0,-0.4){\line(0,1){0.8}}
  \put(0,0){\circle*{0.5}}
  \put(0,1.5){\vector(0,-1){1}}
  \put(-0.2,2){$n_0$}
  \put(-1.5,-1){\footnotesize{$-(N-1)$}}
  \put(1,-0.2){\line(0,1){0.4}}
  \put(2,-0.2){\line(0,1){0.4}}
  \put(3,-0.2){\line(0,1){0.4}}
  \put(4,0){\circle*{0.5}}
  \put(4,1.5){\vector(0,-1){1}}
  \put(2.8,2){$n_0+k$}
  \put(5,-0.2){\line(0,1){0.4}}
  \put(6,-0.2){\line(0,1){0.4}}
  \put(7,-0.2){\line(0,1){0.4}}
  \put(8,0){\circle*{0.5}}
  \put(8,1.5){\vector(0,-1){1}}
  \put(6.8,2){$n_0+2k$}
  \put(8,-0.4){\line(0,1){0.8}}
  \put(7.9,-1){\footnotesize{$0$}}
  \put(9,-0.2){\line(0,1){0.4}}
  \put(10,-0.2){\line(0,1){0.4}}
  \put(11,-0.2){\line(0,1){0.4}}
  \put(12,0){\circle*{0.5}}
  \put(12,-0.2){\line(0,1){0.4}}
  \put(13,-0.2){\line(0,1){0.4}}
  \put(14,-0.2){\line(0,1){0.4}}
  \put(15,-0.2){\line(0,1){0.4}}
  \put(16,-0.4){\line(0,1){0.8}}
  \put(16,0){\circle*{0.5}}
  \put(15.2,-1){\footnotesize{$N-1$}}
  \end{picture} .
\end{center}
Therefore we obtain $k = \frac{N-1}{2}$,
which is a special case of $j=2$.
\\[2mm]
The conditions (i) and (i') do not contradict each other for $N \geq 4$,
and so do not (ii) and (ii') for $N \geq 5$.
Then we obtain $k= \frac{N}{2}$ (resp. $\frac{N-1}{2}$) if $2 | N$ 
(resp. $2 | (N-1))$.

In general $j_{\geq 3}$ cases,
\eqref{condition-0} reduces to
\begin{align}
  \label{condition-j}
  n_0 +(j-1) k < 0, ~~~ 
  \frac{N-n_0}{2 j} \leq k \leq \frac{N-1-n_0}{2 j -1}.
\end{align}
Then two critical cases are written as follows;
\\
(i) when $n_0 = - (N-1)$, \eqref{condition-j} becomes
\begin{equation*}
   \frac{2N-1}{2 j} \leq k \leq \frac{2(N-1)}{2 j -1}.
\end{equation*}
And when $n_0 = -N+k$, $n_0 + (j-1) k = 0$.
Then we obtain $k = \frac{N}{j}$ for $N \geq 2 j$ and $j | N$.
\\
(ii) When $n_0 = - N+k$, 
\begin{equation*}
  \frac{2N}{2 j +1} \leq k \leq \frac{2N-1}{2 j}.
\end{equation*}
And when $n_0 = -(N-1)$, $n_0 + j k = 0$.
Then we get $k = \frac{N-1}{j}$ for $N \geq 2 j +1$ and $j | (N-1)$.
 
Finally we obtain the set $\mathcal{K}$ \eqref{k-K} $k$ belongs to. 
$\square$

\setcounter{equation}{0}

\renewcommand{\theequation}{C.\arabic{equation}}

\addcontentsline{toc}{chapter}{Appendix \protect\numberline{C}
                              Proof of Proposition \ref{prop:subset}}

%%%%%%%%%%%%%%%%%%%%%%%%%%%

\subsection*{Appendix C ~ Proof of Proposition \ref{prop:subset}}

We show the first part of 
Proposition \ref{prop:subset} in more general setting.
Assume that $\mathbf{T}_{LV}(z)$ has a form as 
$$
  \mathbf{T}_{LV}(z) = \boldsymbol{\mu}_0^{LV} z^m + 
                       \boldsymbol{\mu}_1^{LV} z^{m-1}
                  + \cdots + \boldsymbol{\mu}_m^{LV}.
$$ 
Let all matrix elements of $\boldsymbol{\mu}_i^{LV}$
belong to $\mathcal{A}_{LV}^\prime$.
We relate $\mathbf{T}_{LV}(z)$ to a matrix $\mathbf{T}(z)$ 
by the gauge transformation
$$
  \mathbf{T}(z) = \mathbf{B} \mathbf{T}_{LV}(z) \mathbf{B}^{-1}.
$$ 
Here the gauge matrix $\mathbf{B}$ is a diagonal matrix 
independent of $z$,
whose entries belong to a Poisson bracket algebra where 
$\mathcal{A}_{LV}^\prime$ is embedded.
Then the matrix $\mathbf{T}(z)$ has a similar form
to $\mathbf{T}_{LV}(z)$;
$$
  \mathbf{T}(z) = \boldsymbol{\mu}_0 z^m + \boldsymbol{\mu}_1 z^{m-1}
                  + \cdots + \boldsymbol{\mu}_m,
$$
where $\boldsymbol{\mu}_i = 
\mathbf{B} \boldsymbol{\mu}_i^{LV} \mathbf{B}^{-1}$. 
With these settings we have
\\[3mm]
\noindent{\bf Proposition 4.5$\mathrm{^\prime}$}
{\em Let $\mathcal{A}_{N}$ be a Poisson bracket algebra generated by
the entries of a matrix $\mathbf{N}(z)$
related to $\mathbf{T}(z)$ by an invertible matrix $\mathbf{S}$ as
\begin{equation*}
  \mathbf{N}(z) = \mathbf{S} \mathbf{T}(z) \mathbf{S}^{-1},
  ~~
  \mathbf{S} = 
    \begin{pmatrix}
       \vec{e}_i \boldsymbol{\mu}^{(1)} \\
       \vec{e}_i \boldsymbol{\mu}^{(2)} \\
         \vdots                         \\
       \vec{e}_i \boldsymbol{\mu}^{(N)}
    \end{pmatrix}.
\end{equation*}
Here each of $\boldsymbol{\mu}^{(i)}$ is a product 
of $\boldsymbol{\mu}_j$ $(j=0,\cdots, m)$.
Then $\mathcal{A}_N$ is embedded in $\mathcal{A}_{LV}^\prime$.}
\\[3mm]
{\em Proof}.
It is sufficient to show that 
the matrix elements of $\mathbf{N}(z)$ belong to 
$\mathbb{C}(\mathcal{P}_0,V_n;n \in \mathbb{Z}/L\mathbb{Z})$.
Using $\mathbf{B} = \diag[ b_1, b_2, \cdots, b_N ]$,
the matrix $\boldsymbol{\mu}^{(i)}$ is rewritten as
$$
\boldsymbol{\mu}^{(i)} = \mathbf{B} ~ 
                         \boldsymbol{\mu}^{(i)\,LV}~ \mathbf{B}^{-1},
$$
where 
$\boldsymbol{\mu}^{(i)\, LV}$ is the associated 
product of $\boldsymbol{\mu}_j^{LV}$. 
Therefore the gauge matrix $\mathbf{S}$ can be written as
$$
  \mathbf{S} = b_i \mathbf{S}_{LV} \mathbf{B}^{-1},
   ~~~
   \mathbf{S}_{LV} =
   \begin{pmatrix}
     \vec{e}_i \boldsymbol{\mu}^{(1)\, LV} \\
     \vec{e}_i \boldsymbol{\mu}^{(2)\, LV} \\
       \vdots \\
     \vec{e}_i \boldsymbol{\mu}^{(N)\, LV} 
   \end{pmatrix}.
$$
Then $\mathbf{N}(z)$ is obtained as
\begin{align*}
  \mathbf{N}(z) &= 
  b_i \mathbf{S}_{LV} \mathbf{B}^{-1} \mathbf{T}(z) 
  \mathbf{B} \mathbf{S}_{LV}^{-1} b_i^{-1} \\
  &=
  \mathbf{S}_{LV} \mathbf{T}_{LV}(z) \mathbf{S}_{LV}^{-1}.
\end{align*}
Since all entries of $\mathbf{S}_{LV}$ and $\mathbf{T}_{LV}(z)$ 
belong to $\mathcal{A}_{LV}^\prime$,
the proposition follows.  $\square$

When we apply this proposition to
the case $\mathbf{T}(z) = \mathbf{T}_{m;n_1,n_2}(z)$, 
the first part (1) follows. 
Further, from (1) we see that 
the separation equation \eqref{separation}
can be written in terms of entries in
$\mathbf{M}_F(z)$, then we obtain the second part (2).  $\square$
  
%%%%%%%%%%%%%%%%%%%%%%%%%%%%%%%%%%
%%%%%%%%%%%%%%%%%%%%%%%%%%%%%%%%%%


\begin{thebibliography}{10}

\bibitem{DubMatNov76}
B.~A. Dubrovin, V.~B. Matveev, and S.~P. Novilov, Uspekhi Mat. Nauk
  \textbf{31}, 55 (1976).

\bibitem{Krichever78}
I.~M. Krichever, Uspekhi Mat. Nauk \textbf{34}, 215 (1978).

\bibitem{MoerMum79}
P.~van Moerbeke and D.~Mumford, Acta Math. \textbf{143}, 93 (1979).

\bibitem{AdlerMoer80}
M.~Adler and P.~van Moerbeke, Adv. in Math. \textbf{38}, 267 (1980),
Adv. in Math. \textbf{38}, 318 (1980).

\bibitem{ReySemenov94}
A.~G. Reyman and M.~A. Semenov-Tian-Shansky, {\it Dynamical Systems VII\/},
  vol. \textbf{16} of {\it Encyclopedia of Mathematical Sciences\/}, 116--225
  (Springer-Verlag, Berlin, 1994).

\bibitem{Griffiths85}
P.~A. Griffiths, Amer. J. Math. \textbf{107}, 1445 (1985).

\bibitem{Beauville90}
A.~Beauville, Acta. Math. \textbf{164}, 211 (1990).

\bibitem{Smirnov-Zeitlin0203}
F.~A. Smirnov and V.~Zeitlin, math-ph/0203037.

\bibitem{Mumford-Book}
D.~Mumford, {\it Tata Lectures on Theta II\/} (Birkh\"auser, 1984).

\bibitem{DonagiMarkman96}
R.~Donagi and E.~Markman, Lecture Notes in Mathematics \textbf{1620}, 1 (1996).

\bibitem{Medan99}
C.~M\'edan, Math. Z. \textbf{232}, 665 (1999).

\bibitem{SmirnovNakayashiki00}
A.~Nakayashiki and F.~A. Smirnov, Comm. Math. Phys. \textbf{217}, 623 (2001),
  math-ph/0001017.

\bibitem{Van1638}
P.~Vanhaecke, Lecture Notes in Mathematics \textbf{1638} (2001).

\bibitem{Vivolo03}
O.~Vivolo, J. Geom. Phys. \textbf{46}, 99 (2003).

\bibitem{Smirnov-Zeitlin0111}
F.~A. Smirnov and V.~Zeitlin, math-ph/0111038.

\bibitem{Sklyanin95}
E.~K. Sklyanin, Prog. Theor. Phys. Suppl. \textbf{118}, 35 (1995).

\bibitem{Bogo88}
O.~I. Bogoyavlensky, Phys. Lett. A \textbf{134}, 34 (1988).

\bibitem{Suris94}
Y.~B. Suris, Phys. Lett. A \textbf{188}, 256 (1994).

\bibitem{InoueHikami98-Bogo}
R.~Inoue and K.~Hikami, J. Phys. Soc. Jpn. \textbf{67}, 87 (1998).

\bibitem{FadTak86}
L.~D. Faddeev and L.~A. Takhtajan, Lect. Notes Phys. \textbf{246}, 166 (1986).

\bibitem{BonoraColCon96}
L.~Bonora, L.~P. Colatto, and C.~P. Constantinidis, Phys. Lett. B \textbf{387},
  759 (1996).

\bibitem{AntoBelovChal97}
A.~V. Antonov, A.~A. Belov, and K.~D. Chaltikian, J. Geom. Phys. 22, 298
  (1997).

\bibitem{FrenkelReshSemenov97}
E.~Frenkel, N.~Reshetikhin, and M.~A. Semenov-Tian-Shansky, Comm. Math. Phys.
  \textbf{192}, 605 (1998).

\bibitem{HikamiSogoInoue97}
K.~Hikami, K.~Sogo, and R.~Inoue, J. Phys. Soc. Jpn. \textbf{66}, 3756 (1997).

\bibitem{Vanhae01}
R.~L. Fernandes and P.~Vanhaecke, Comm. Math. Phys. \textbf{221}, 169 (2001).

\bibitem{Inoue02-JBogo}
R.~Inoue, J. Math. Phys. \textbf{44}, 338 (2003).

\bibitem{AdlerMoer89}
M.~Adler and P.~van Moerbeke, Invent. Math. \textbf{97}, 3 (1989).

\bibitem{FadTak87}
L.~D. Faddeev and L.~A. Takhtajan, {\it Hamiltonian methods in the theory of
  solitons\/} (Springer-Verlag, Berlin, 1987).

\bibitem{Sklyanin85}
E.~K. Sklyanin, Lecture notes in Physics \textbf{226}, 196 (1985).

\bibitem{Sklyanin92}
E.~K. Sklyanin, Comm. Math. Phys. \textbf{150}, 181 (1992).

\bibitem{Scott94}
D.~R. Scott, J. Math. Phys. \text{35}, 5831 (1994), hep-th/9403030.

\bibitem{Gekhtman95}
M.~I. Gekhtman, Comm. Math. Phys. \textbf{167}, 593 (1995).

\bibitem{Griffiths-Book}
P.~A. Griffiths, {\it Introduction to algebraic curves\/} (AMS, 1989).

\bibitem{Smirnov0001}
F.~A. Smirnov, J. Phys. A: Math. Gen. \textbf{33}, 3385 (2000),
  math-ph/0001032.

\bibitem{HikamiInoueKomori99}
K.~Hikami, R.~Inoue, and Y.~Komori, J. Phys. Soc. Jpn. \textbf{68}, 2234
  (1999).

\end{thebibliography}
\end{document}